\documentclass[%
preprint,
 amsmath,amssymb,
 aps,
pra,
]{revtex4-1}

\usepackage{graphicx}
\usepackage{dcolumn}
\usepackage{bm}

\usepackage{color}
\usepackage{float}

\begin{document}

\title{Conical Intersections Induced by Quantum Light: \\ Field-Dressed Spectra
from the Weak to the Ultrastrong Coupling Regimes}

\author{Tam\'as Szidarovszky}
    \email{tamas821@caesar.elte.hu}
    \affiliation{Laboratory of Molecular Structure and Dynamics, Institute of Chemistry, E\"otv\"os Lor\'and University and MTA-ELTE Complex Chemical Systems Research Group, 
	H-1117 Budapest, P\'azm\'any P\'eter s\'et\'any 1/A, Hungary}%

\author{G\'abor J. Hal\'asz}
	\affiliation{Department of Information Technology, University of Debrecen, P.O. Box 400, H-4002 Debrecen, Hungary}

\author{Attila G. Cs\'asz\'ar}
	\affiliation{Laboratory of Molecular Structure and Dynamics, Institute of Chemistry, E\"otv\"os Lor\'and University and MTA-ELTE Complex Chemical Systems Research Group,
	H-1117 Budapest, P\'azm\'any P\'eter s\'et\'any 1/A, Hungary}

\author{Lorenz S. Cederbaum}
	
	\affiliation{Theoretische Chemie, Physikalisch-Chemisches Institut, Universit\"at Heidelberg, D-69120, Heidelberg, Germany}	

\author{\'Agnes Vib\'ok}
    \email{vibok@phys.unideb.hu}
	\affiliation{Department of Theoretical Physics, University of Debrecen, P.O. Box 400, H-4002 Debrecen, Hungary and ELI-ALPS, ELI-HU Non-Profit Ltd., Dugonics t\'er 13, H-6720 Szeged, Hungary}

\date{\today}


\begin{abstract}
    A fundamental theoretical framework is formulated for the investigation of rovibronic spectra resulting from the coupling of molecules to one mode of the radiation field in an optical cavity. The approach involves the computation of (1) cavity-field-dressed rovibronic states, which are hybrid light-matter eigenstates of the ``molecule + cavity radiation field'' system, and (2) the transition amplitudes between these field-dressed states with respect to a weak probe pulse. The predictions of the theory are shown for the homonuclear Na$_2$ molecule.
    The field-dressed rovibronic spectrum demonstrates undoubtedly that the Born--Oppenheimer approximation breaks down in the presence of the cavity radiation field. A clear fingerprint of the strong nonadiabaticity is found, which can only emerge in the close vicinity of conical intersections. In this work, the conical intersection is induced by the quantized radiation field, and it is thus called a "light-induced conical intersection" (LICI). Dependence of the cavity-field-dressed spectrum on the cavity-mode wavelength as well as on the light-matter coupling strength is investigated. Essential changes are identified in the spectra from the weak to the ultrastrong coupling regimes.
\end{abstract}

\maketitle

\section{\label{Introduction}Introduction}

    Understanding the interaction of matter with strong and ultrastrong laser pulses is a fundamental and rapidly developing field of research. With the remarkable advances
    in laser technology in the past few decades, experimental investigation 
    of the interaction became feasible \cite{fs_lasers_RevModPhys_2000,as_lasers_RevModPhys_2009,strong_fields_in_periodic_systems_RevModPhys_2018} and provided insight into the strongly nonlinear
    domain of optical processes, associated with various unique phenomena, such
    as high harmonic generation \cite{HHG_Kulander_PRL_1992,HHG_theory_Corkum_PRA_1994}, above threshold
    dissociation and ionization \cite{ATI_Kulander_PRL_1993}, bond softening and
    hardening effects \cite{Bandrauk1,Bandrauk_JPC_1987,Bandrauk2,Bandrauk3,Bucksbaum1,Bucksbaum2,Bucksbaum3},
    and light-induced conical intersections (LICIs) \cite{LICI1,LICI2,LICI3,LICI4,LICI5,Gabi6,Bandrauk_LICI,Nimrod2,LICI7}\footnote{Conical intersections (CIs) are geometries where two electronic states of a molecule share the same energy, providing a very efficient channel for nonradiative relaxation processes to the ground state on an ultrafast time scale. For CIs to be formed in a molecular system one needs two independent degrees of freedom, which constitute the space in which the CIs can exist. Therefore, CIs between different electronic states can only occur for molecules with at least three atoms. For a diatomic molecule, which has only one vibrational degree of freedom, it is not possible for two electronic states of the same symmetry to become degenerate, as required by the well-known noncrossing rule. However, this statement is true only in free space.}.
    
    LICIs may form even in diatomic molecules, when the laser light
    not only rotates the molecule but can also couple the vibrational with the emerging rotational degree of freedom. Theoretical and experimental studies have demonstrated that the light-induced nonadiabatic effects have significant impact on different observable dynamical properties, such as molecular alignment, dissociation probability, or angular distribution of photofragments \cite{LICI1,LICI2,LICI3,LICI4,LICI5,Gabi6,Bandrauk_LICI}.
    Recently, signatures of light-induced nonadiabatic phenomena
    have been successfully identified in the classical field-dressed static rovibronic spectrum of diatomics \cite{LICI_in_spectrum_Szidarovszky_JPCL_2018}.
    
    As an alternative to interactions of atoms or molecules with intense laser fields, strong
    light-matter coupling can also be achieved, both for atoms and molecules, by their confinement
    in microscale or nanoscale optical cavities \cite{Edina, Ruggenthaler2018, Domokos_PRA_2015}. Such systems are usually described in terms of field-dressed or polariton states, which are the eigenstates of the full ``atom/molecule + radiation field'' system \cite{cavity_Ebbesen_AccChemRes_2016,cavity_Flick_PNAS_2017,cavity_Feist_ACSphotonics_2018, review_ultrastrong_coupling_Kockum_2018}. 
    With decreasing cavity size the quantized nature of the radiation field eventually becomes important
    and strong photon-matter coupling as well as a significant modification
    of the atomic and molecular properties may occur even if the photon
    number is (close to) zero \cite{cavity_Ebbesen_AccChemRes_2016,cavity_Flick_PNAS_2017,cavity_Feist_ACSphotonics_2018,cavity_Herrera_PRL_2016,cavity_Galego_NatCommun_2016,cavity_Galego_PRL_2017,cavity_Luk_JCTC_2017}.
    For example, quantum modeling efforts have shown that strong resonant coupling of a cavity radiation
    field with an electronic transition can decouple the electronic and
    nuclear degrees of freedom in molecular ensembles \cite{cavity_Herrera_PRL_2016},
    strong coupling of molecules to a confined light mode could suppress
    photoisomerization \cite{cavity_Galego_NatCommun_2016}, and that collective motion of
    molecules could be triggered by a single photon of a cavity \cite{cavity_Galego_PRL_2017,cavity_Luk_JCTC_2017,collective_CI_cavity_Vendrell_arxiv_2018}.
    In addition to exploring strong light-matter coupling, processes in
    nanoscale cavities could also be used to study how atoms/molecules interact
    with non-classical states of the quantized light field \cite{cavity_Triana_JPCA_2018,cavity_MCTDH_Vendrell_CP_2018}.
    
    Previous theoretical models mostly treated atoms or molecules in reduced
    dimensions or via some simplified model \cite{cavity_Herrera_PRL_2016,cavity_Galego_NatCommun_2016,cavity_Triana_JPCA_2018,cavity_MCTDH_Vendrell_CP_2018,cavity_Galego_PRX_2015,cavity_Kowalewski_JPCL_2016, cavity_chemical_reactivity_Feist_arxiv_2018,spectra_of_vibDressedpolaritons_Zeb_ACSphotonics_2018,dark_vibronic_polaritons_Herrera_PRL_2017,theory_of_organic_cavities_Herrera_ACSphotonics_2018},
    and most often utilized the concept of polariton states formed by
    coupling the electronic and photonic degrees of freedom. Nuclear motion
    is then thought to proceed on the polariton surfaces. Alternatively,
    decoupling the electronic motion from the nuclear and photonic degrees
    of freedom, known as the cavity Born--Oppenheimer approximation, has
    sometimes been pursued \cite{cavity_Flick_PNAS_2017,cavityBO_Flick_JCTC_2017}.
    
    On the experimental side, the coupling of molecules to cavity radiation
    field was shown to modify chemical landscapes and reaction dynamics
    \cite{cavity_exp_Hutchison_AngChem_2012}, as well as the absorption spectra \cite{cavity_Schwartz_CPC_2013,ultrastrong_coupling_spectrosc_Jino_FarDisc_2015}
    of molecules. Furthermore, intermolecular non-radiative energy transfer
    was found to be enhanced by the formation of polariton states \cite{cavity_Zhong_AngChem_2016},
    and the hybridization of molecular vibrational states through strong light-matter
    coupling in a microcavity \cite{cavity_Muallem_JPCL_2016} was also observed.
    
\begin{figure}[h]
    \includegraphics[width=0.5\textwidth]{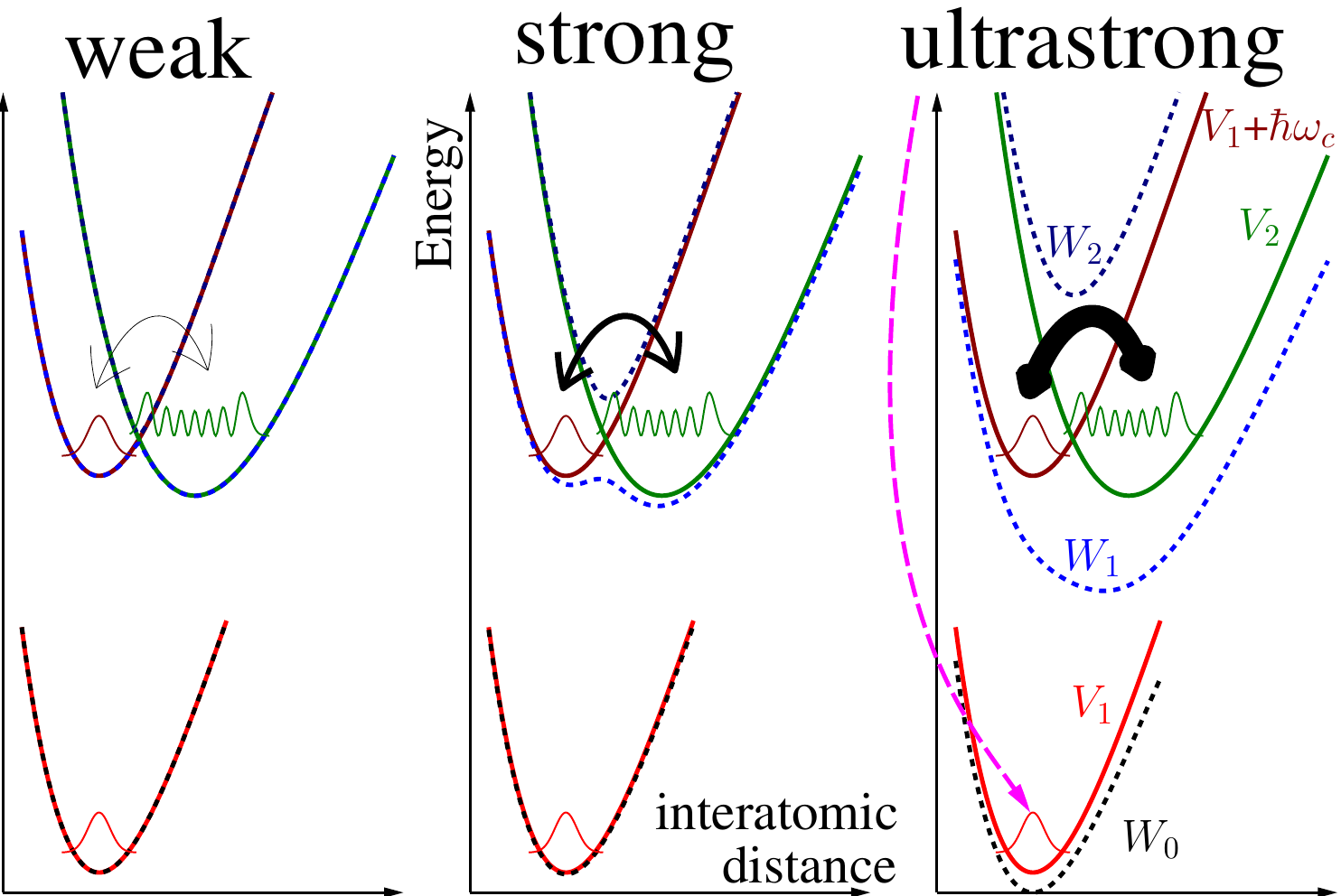}
    \caption{\label{Fig:couplingRegions}The three regimes of coupling strength , and the related field-dressed PECs, of a molecule interacting with a resonant cavity mode. The diabatic surfaces $V_1$ and $V_2$ are indicated with continuous line, while the polariton surfaces $W_0, W_1$ and $W_2$ are indicated with dashed lines. $\hslash \omega_c$ is the cavity photon energy.}
\end{figure}

    For atoms or molecules interacting with a cavity mode (near) resonant to an electronic transition, one usually distinguishes three regimes of field-matter coupling strengths \cite{cavity_Flick_PNAS_2017,cavity_Galego_PRX_2015}, as depicted in Fig. \ref{Fig:couplingRegions}: weak, strong, and ultrastrong. In the weak-coupling regime, see the left panel of Fig. \ref{Fig:couplingRegions}, the diabatic picture of photon-dressed potential energy curves (PECs) holds and the cavity mode only couples the excited electronic state with the ground electronic state dressed by a photon. In the strong-coupling regime, shown in the middle panel of Fig. \ref{Fig:couplingRegions}, polariton states are formed and the adiabatic picture becomes appropriate for describing the excited state manifold, while the ground state remains essentially unchanged. Finally, in the ultrastrong-coupling regime, see the right panel of Fig. \ref{Fig:couplingRegions}, nonresonant couplings become strong enough to significantly modify the electronic ground state, as well.

\begin{figure}[h]
    \includegraphics[width=0.5\textwidth]{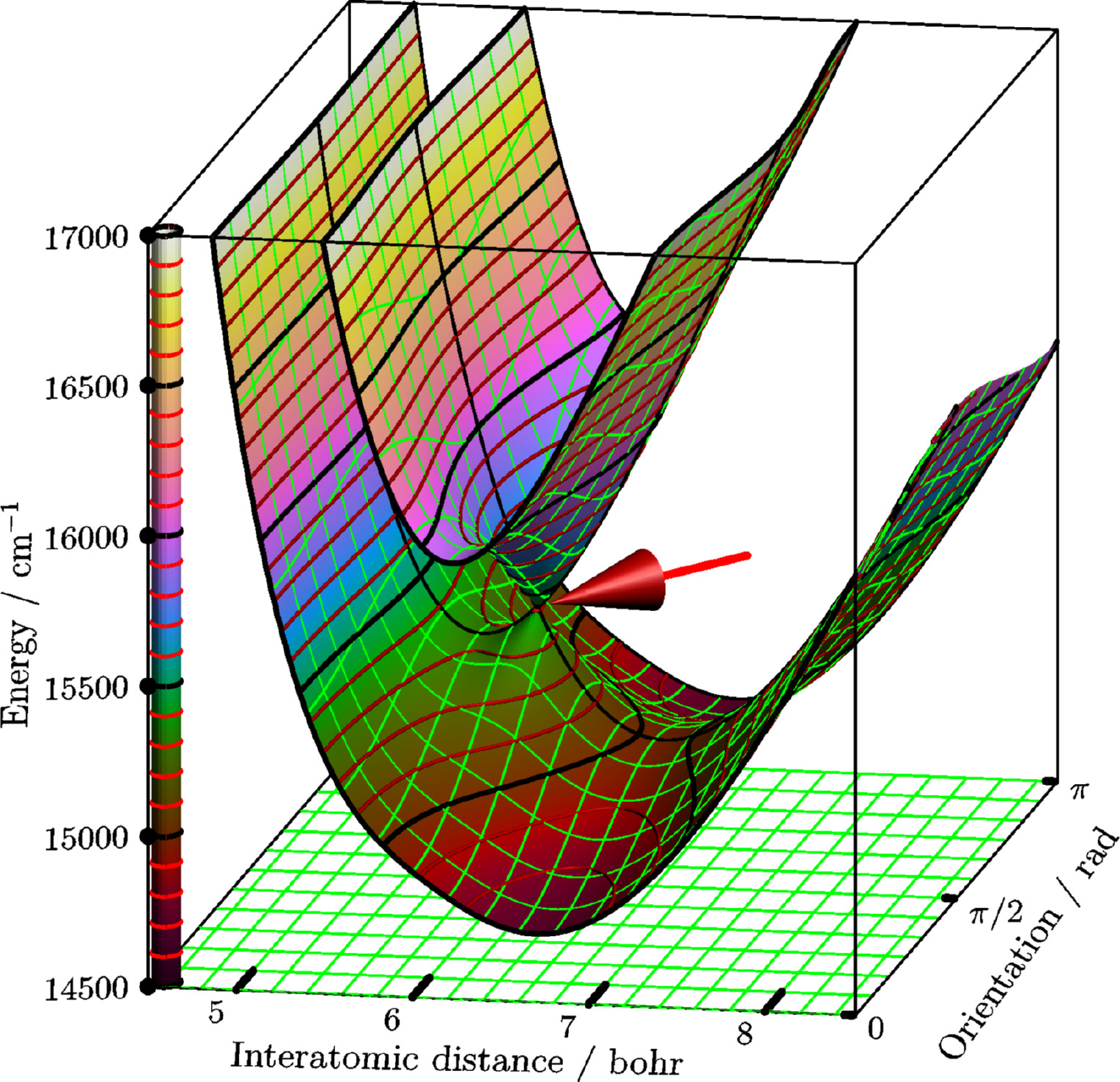}
    \caption{\label{Fig:LICI}Two dimensional polariton surfaces  ($W_1$ and $W_2$) of  the Na$_2$ dimer in the cavity. The one photon coupling of the cavity ($\varepsilon_c$) corresponds to a classical field intensity of 64 GWcm$^{-2}$. The cavity mode wavelength is $\lambda_c=653$ nm. The red arrow denotes the position of the light-induced conical intersection.} 
\end{figure}

    In reality, the dashed polariton surfaces of Fig. \ref{Fig:couplingRegions} are strictly valid only if the molecular axis is parallel to the preferred polarization direction of the cavity field. 
    In contrast, when the molecular axis is perpendicular to the polarization direction of the cavity, the light-matter coupling vanishes and the diabatic picture (continuous potentials in Fig. \ref{Fig:couplingRegions}) is the relevant one. 
    In fact, the orientation of a rotating molecule can change continuously between these two extreme positions, and the diabatic and cavity-induced polariton surfaces are continuously transformed into each other.
    Therefore, due to the rotation of the molecule, the upper and lower adiabatic surfaces are not completely separated but a conical intersection (CI) emerges between them, see Fig. \ref{Fig:LICI}, at which point the nonadiabatic couplings become infinitely strong. 

    At the vicinity of this CI, which is created by the quantum light and never present in field-free diatomics, the Born$-$Oppenheimer picture \cite{27BoOp,54BoHu} breaks down. The nuclear dynamics proceed on the coupled polariton surfaces and motions along the vibrational and rotational coordinates become intricately coupled. It must be stressed that even in the case of diatomics, considering rotations completely changes the paradigm and physical picture with respect to the description when only the vibrational, electronic, and photonic degrees of freedom are taken into account.
    Moreover, in contrast to field-free polyatomic molecules, where conical intersections are dictated by nature, they are either present or not, light-induced conical intersections in the cavity are always present between the polariton surfaces. Even for a diatomic molecule, the appropriate description needs to account for rotations, which are coupled nonadiabatically to the vibrational, electronic, and photonic modes of the system.

    The purpose of the present study
    is to investigate the field-dressed rovibronic spectrum of diatomics
    in the framework of cavity quantum electrodynamics (QED). We complement previous theoretical approaches by accounting for all molecular degrees of freedom, \textit{i.e.}, we treat rotational, vibrational, electronic, and photonic degrees of freedom on an equal footing. Furthermore, we incorporate for the first time the concept of LICIs with the quantized radiation field.
    
    Our goals with are two-fold.
    First, we investigate the field-dressed rovibronic spectrum of our test system, the homonuclear Na$_2$ molecule, in order to understand the effects of the cavity on the spectrum and to identify the direct signatures of a LICI created by the quantized cavity radiation field.
    Second, the coupling strength and cavity-mode wavelength dependence of the spectrum from weak to ultrastrong coupling regimes is investigated. We identify the formation of polariton states in the strong coupling regime as well as the impact of nonresonant couplings on the spectrum in the ultrastrong coupling regime. Surprisingly, the effects of nonresonant couplings can be seen at coupling strengths much smaller than those necessary for significantly distorting the ground-state PEC, as long as molecular rotations are properly accounted for. An important conclusion of this study is that for the simulation of freely rotating molecules confined in a cavity the appropriate treatment of rotations as well as vibrations is mandatory.

\section{\label{Theory}Theoretical approach}

    For simulating the weak-field absorption spectrum of molecules confined in small optical cavities, we first
    determine the field-dressed states, \textit{i.e.}, the eigenstates
    of the full ``molecule + radiation field'' system, and then we compute the dipole transition amplitudes
    between the field-dressed states with respect to a probe pulse. We
    assume the probe pulse to be weak; therefore, transitions induced
    by it should be dominated by one-photon processes. This implies that
    the standard approach \cite{BunkerJensen} of using first-order time-dependent
    perturbation theory to compute the transition amplitudes should be
    adequate.

\subsection{The field-dressed states} 

    Within the framework of QED and the electric dipole representation, the Hamiltonian of a molecule interacting with a single cavity mode can be written as \cite{Cohen-Tannoudji}
    \begin{equation}
    \hat{H}_{{\rm tot}}=\hat{H}_{{\rm mol}}+\hat{H}_{{\rm rad}}+\hat{H}_{{\rm int}},\label{eq:CavityHamiltonian_general}
    \end{equation}
    where $\hat{H}_{{\rm mol}}$ is the field-free molecular Hamiltonian,
    $\hat{H}_{{\rm rad}}$ is the radiation-field Hamiltonian, and $\hat{H}_{{\rm int}}$
    is the interaction term between the molecular dipole moment and the
    electric field. For a single radiation mode and an appropriate choice
    of the origin \cite{Cohen-Tannoudji},
    \begin{equation}
    \hat{H}_{{\rm rad}}=\hslash\omega _c\hat{a}^{\dagger}\hat{a}\label{eq:Hrad}
    \end{equation}
    and 
    \begin{equation}
    \hat{H}_{{\rm int}}=-\sqrt{\frac{\hslash\omega _c}{2\epsilon_{0}V}}\mathbf{\hat{d}\hat{e}}\left(\hat{a}^{\dagger}+\hat{a}\right)
    =-\frac{\varepsilon_c}{\sqrt{2}}\mathbf{\hat{d}\hat{e}}\left(\hat{a}^{\dagger}+\hat{a}\right),\label{eq:Hint}
    \end{equation}
    where $\varepsilon_c = \sqrt{\hslash\omega _c/(\epsilon_{0}V)}$  is the cavity one-photon field, $\hat{a}^{\dagger}$ and $\hat{a}$ are photon creation
    and annihilation operators, respectively, $\omega_c$ is the frequency
    of the cavity mode, $\hslash$ is Planck's constant divided by $2\pi$,
    $\epsilon_{0}$ is the electric constant, $V$ is the volume of the
    electromagnetic mode, $\mathbf{\hat{d}}$ is the molecular dipole
    moment, and $\mathbf{\hat{e}}$ is the polarization vector of the
    cavity mode.
    
    In the case of diatomic molecules, representing the Hamiltonian of
    Eq. (\ref{eq:CavityHamiltonian_general}) in a direct product basis
    composed of two field-free molecular electronic states and the Fock
    states of the radiation field gives
    
    \begin{equation}
    \hat{H}=\begin{bmatrix}\hat{H}_{{\rm m}} & \hat{A}(2) & 0 & \ldots\\
    \hat{A}^{\dagger}(2) & \hat{H}_{{\rm m}}+\hslash\omega_c & \hat{A}(3) & \ldots\\
    0 & \hat{A}^{\dagger}(3) & \hat{H}_{{\rm m}}+2\hslash\omega_c & \ldots\\
    \vdots & \vdots & \vdots & \ddots
    \end{bmatrix},\label{eq:CavityHamiltonian}
    \end{equation}
    where 
    \begin{equation}
    \hat{H}_{{\rm m}}=\begin{bmatrix}\hat{T} & 0\\
    0 & \hat{T}
    \end{bmatrix}+\begin{bmatrix}V_{1}(R) & 0\\
    0 & V_{2}(R)
    \end{bmatrix},\label{eq:MolecularHamiltonian}
    \end{equation}
    and 
    \begin{equation}
    \hat{A}(N)=\begin{bmatrix}g_{11}(R,\theta)\sqrt{N} & g_{12}(R,\theta)\sqrt{N}\\
    g_{21}(R,\theta)\sqrt{N} & g_{22}(R,\theta)\sqrt{N}
    \end{bmatrix},\label{eq:CavityAmx}
    \end{equation}
    with 
    \begin{equation}
    g_{ij}(R,\theta)=-\sqrt{\frac{\hslash\omega}{2\epsilon_{0}V}}d_{ij}(R){\rm cos}(\theta),\label{eq:quantized_A}
    \end{equation}
    where $R$ is the internuclear distance, $V_{i}(R)$ is the $i$th
    PEC, $\hat{T}$ is the nuclear kinetic energy
    operator, $d_{ij}(R)$ is the transition dipole moment matrix element
    between the $i$th and $j$th electronic states, and $\theta$ is
    the angle between the electric field polarization vector and the transition
    dipole vector, assumed to be parallel to the molecular axis.
    
    In Eq. (\ref{eq:CavityHamiltonian}), the first, second, third, etc.
    columns(rows) correspond to zero, one, two, etc. photon number in
    the bra(ket) vectors of the cavity mode, respectively.
    Expanding Eq. (\ref{eq:CavityHamiltonian}) using Eqs. (\ref{eq:MolecularHamiltonian})
    and (\ref{eq:CavityAmx}), and assuming a homonuclear diatomic molecule having no permanent dipole, gives 
    \begin{equation}
    \hat{H}=
    \begin{bmatrix}\hat{T}+V_{1}(R) & 0 & 0 & g_{12}(R,\theta)\sqrt{2} & 0 & \cdots \\
    0 & \hat{T}+V_{2}(R) & g_{21}(R,\theta)\sqrt{2} & 0 & 0 & \cdots \\
    0 & g_{12}(R,\theta)\sqrt{2} & \hat{T}+V_{1}(R)+\hslash\omega_c & 0 & 0 & \cdots \\
    g_{21}(R,\theta)\sqrt{2} & 0 & 0 & \hat{T}+V_{2}(R)+\hslash\omega_c & g_{21}(R,\theta)\sqrt{3} & \cdots \\
    0 & 0 & 0 & g_{12}(R,\theta)\sqrt{3} & \hat{T}+V_{1}(R)+2\hslash\omega_c & \cdots\\
    \vdots & \vdots & \vdots & \vdots & \vdots & \ddots
    \end{bmatrix},\label{eq:CavityHamiltonian_detailed}
    \end{equation}
    which is the working Hamiltonian used in this study.
    
    The $\vert\Psi_{i}^{{\rm FD}}\rangle$ field-dressed states, \textit{i.e.},
    the eigenstates of the Hamiltonian of Eq. (\ref{eq:CavityHamiltonian_general}),
    \begin{equation}
    \hat{H}_{{\rm tot}}\vert\Psi_{i}^{{\rm FD}}\rangle=E_{i}^{{\rm FD}}\vert\Psi_{i}^{{\rm FD}}\rangle
    \end{equation}
    are obtained by diagonalizing the Hamiltonian of Eq. (\ref{eq:CavityHamiltonian_detailed})
    in the basis of field-free rovibrational states. Then, the field-dressed states can be expressed as the linear
    combination of products of field-free molecular rovibronic states
    and Fock states of the dressing field, $i.e.$, 
    \begin{equation}
    \vert\Psi_{i}^{{\rm FD}}\rangle=\sum_{J,v,\alpha,N}C_{i,\alpha vJN}\vert\alpha vJ\rangle\vert N\rangle=\sum_{J,v,N}C_{i,1vJN}\vert1vJ\rangle\vert N\rangle+\sum_{J,v,N}C_{i,2vJN}\vert2vJ\rangle\vert N\rangle,\label{eq:FieldDressedStates_small_photon_number}
    \end{equation}
    where $\vert jvJ\rangle$ is a field-free rovibronic state, in which
    the molecule is in the $j$th electronic, $v$th vibrational, and
    $J$th rotational state, $\vert N\rangle$ is a Fock state of the
    dressing field with photon number $N$, and $C_{i,jvJN}$ are
    expansion coefficients obtained by diagonalizing the Hamiltonian of
    Eq. (\ref{eq:CavityHamiltonian_detailed}) in the basis of the field-free
    rovibrational states.

\subsection{Transitions between field-dressed states}

    Let us now compute the absorption spectrum with respect to a weak probe
    pulse, whose photon number is represented by the letter $M$. Using
    first-order time-dependent perturbation theory, the transition amplitude
    between two field-dressed states, induced by the weak probe pulse,
    can be expressed as \cite{Jaynes_Cummings_1963,Cohen-Tannoudji}
    \begin{equation}
    \langle\Psi_{i}^{{\rm FD}}\vert\langle M\vert\mathbf{\hat{d}\hat{E}}\vert M'\rangle\vert\Psi_{j}^{{\rm FD}}\rangle=\langle\Psi_{i}^{{\rm FD}}\vert\hat{d}{\rm cos}(\theta)\vert\Psi_{j}^{{\rm FD}}\rangle\langle M\vert\hat{E}\vert M'\rangle.\label{eq:transition_amplitude_general}
    \end{equation}
    In Eq. (\ref{eq:transition_amplitude_general}), the electric field
    operator $\hat{E}$ stands for the weak probe pulse, and we
    assume that the probe pulse has a polarization axis identical to
    that of the cavity mode. Since $\hat{E}$ is proportional to the sum
    of a creation and an annihilation operator acting on $\vert M'\rangle$,
    Eq. (\ref{eq:transition_amplitude_general}) leads to the well-known
    result that the transition amplitude is non-zero only if $M=M'\pm1$,
    $i.e.$, Eq. (\ref{eq:transition_amplitude_general}) accounts for
    single-photon absorption or stimulated emission. 
    
    The matrix element of the operator $\hat{d}{\rm cos}(\theta)$ between
    two field-dressed states of Eq. (\ref{eq:FieldDressedStates_small_photon_number})
    gives 
    \begin{equation}
    \begin{aligned} & \langle\Psi_{i}^{{\rm FD}}\vert\hat{d}{\rm cos}(\theta)\vert\Psi_{j}^{{\rm FD}}\rangle=\\
     & =\left(\sum_{J,v,\alpha,N}C_{i,\alpha vJN}^{*}\langle\alpha vJ\vert\langle N\vert\right)\hat{d}{\rm cos}(\theta)\left(\sum_{J',v',\alpha',N'}C_{j,\alpha'v'J'N'}\vert\alpha'v'J'\rangle\vert N'\rangle\right)=\\
     & =\sum_{J,v,\alpha,N,J',v',\alpha',N'}C_{i,\alpha vJN}^{*}C_{j,\alpha'v'J'N'}\langle\alpha vJ\vert\hat{d}{\rm cos}(\theta)\vert\alpha'v'J'\rangle\delta_{N,N'}=\\
     & =\sum_{J,v,J',v',N}C_{i,1vJN}^{*}C_{j,2v'J'N}\langle1vJ\vert\hat{d}{\rm cos}(\theta)\vert2v'J'\rangle+\sum_{J,v,J',v',N}C_{i,2vJN}^{*}C_{j,1v'J'N}\langle2vJ\vert\hat{d}{\rm cos}(\theta)\vert1v'J'\rangle.
    \end{aligned}
    \label{eq:transition_amplitude_between_cavity_FD_states}
    \end{equation}
    In the last line of Eq. (\ref{eq:transition_amplitude_between_cavity_FD_states}),
    the first(second) term represents transitions, in which the first(second) electronic
    state contributes from the $i$th field-dressed state and the second(first)
    electronic state contributes from the $j$th field-dressed state.
    Assuming that the \textit{i}th state is the initial state, the first term
    in the last line of Eq. (\ref{eq:transition_amplitude_between_cavity_FD_states})
    leads to the usual field-free absorption spectrum in the limit of
    the light-matter coupling going to zero. In all spectra shown
    below, we plot the absolute square of the transition amplitudes, as
    computed by Eq. (\ref{eq:transition_amplitude_between_cavity_FD_states}),
    or their convolution with a Gaussian function.

\section{Computational details}

    We test the theoretical framework developed in Eqs. (\ref{eq:CavityHamiltonian_general}-\ref{eq:transition_amplitude_between_cavity_FD_states}) on the Na$_{2}$ molecule, for which the $V_{1}(R)$ and $V_{2}(R)$ PECs correspond to the ${\rm X}^{1}\Sigma{\rm _{g}^{+}}$ and the 
    ${\rm A}^{1}\Sigma{\rm _{u}^{+}}$ electronic states, respectively.
    The PECs and the transition dipole are taken
    from Refs. \citenum{Na2_PEC_Magnier_JCP_1993} and \citenum{Na2_TDM_Zemke_JMS_1981}, respectively. The field-free rovibrational eigenstates of Na$_2$ on the $V_{1}(R)$ and $V_{2}(R)$ PECs are computed using 200 spherical-DVR basis function \cite{D2FOPI_Szidarovszky_PCCP_2010} with the related grid points placed in the internuclear coordinate range $(0,10)$ bohr. 
    Unless indicated otherwise, the set of field-free rovibrational eigenstates used to represent the Hamiltonian of Eq. (\ref{eq:CavityHamiltonian_detailed}) was composed of all states with $J<30$ and an energy not exceeding the zero point energy of the respective PEC by more than 5000 cm$^{-1}$. The maximum photon number in the cavity mode was set to two.  
    
\section{Results and discussion}

\begin{figure}[h]
    \begin{minipage}{0.58\columnwidth}%
     \includegraphics[width=0.95\textwidth]{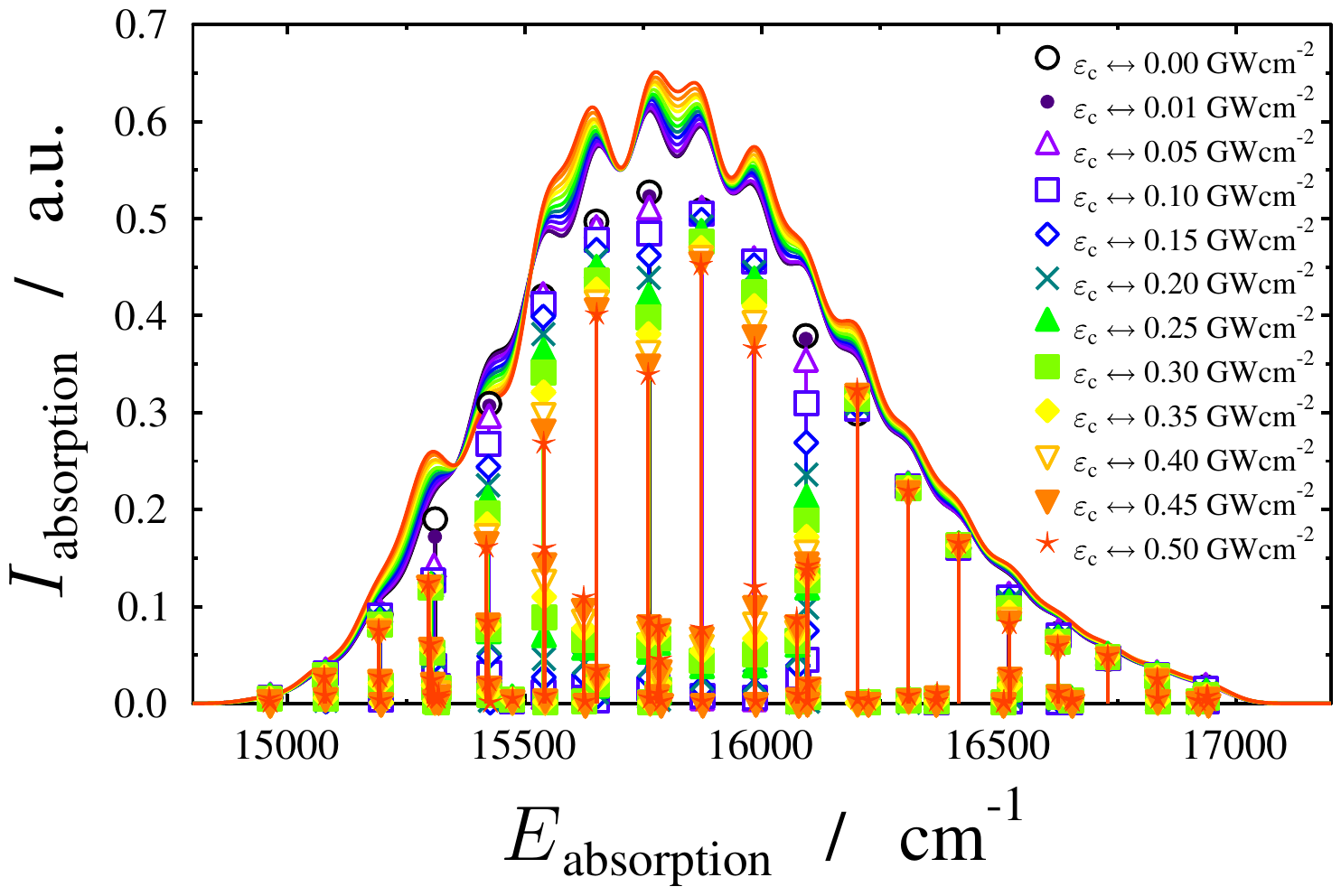}
    \end{minipage}
    \begin{minipage}{0.41\columnwidth}%
        \includegraphics[width=0.95\textwidth]{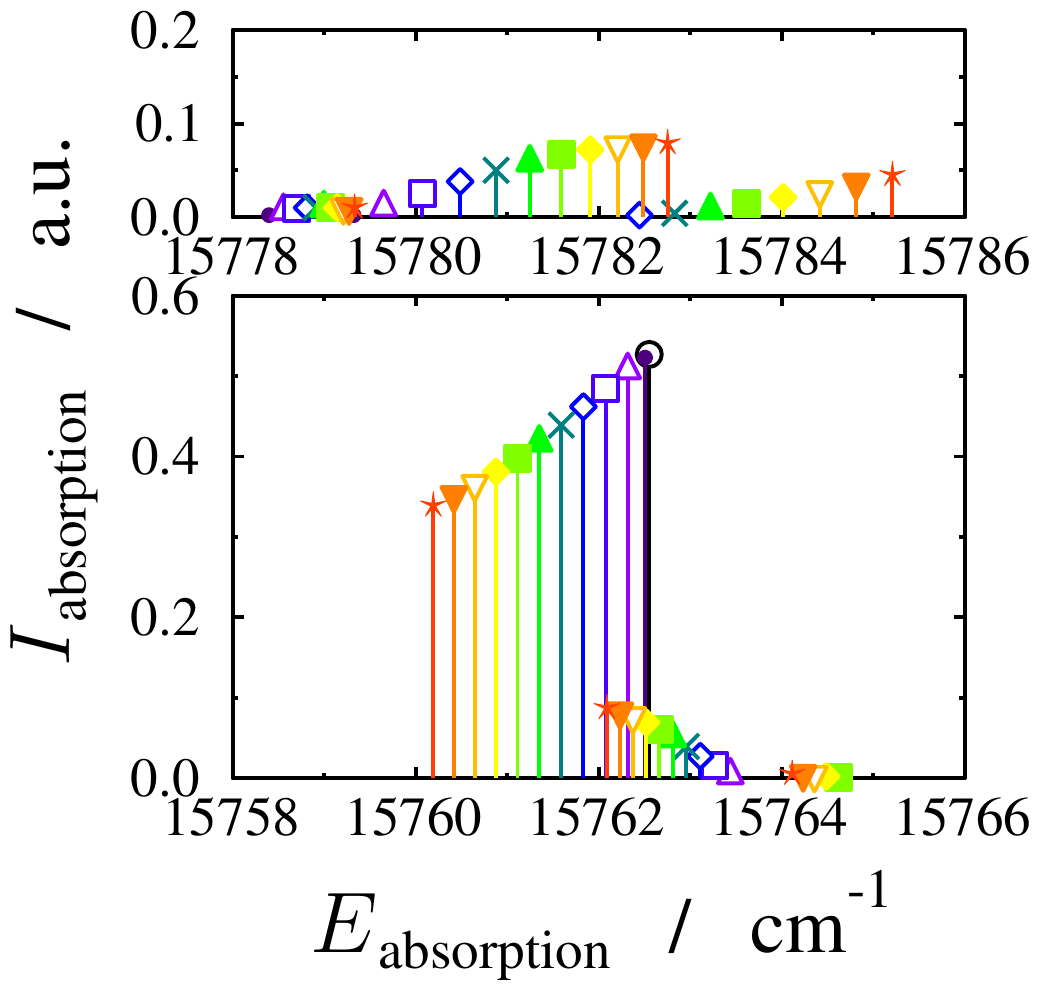}
    \end{minipage}
    
    \caption{\label{FDS_weakcoupling}Field-dressed spectra obtained with different values of the light-matter coupling strengths for a cavity mode wavelength of $\lambda=653$ nm. Coupling strength values are indicated by the intensity of a classical light field giving a coupling strength equal to the one-photon coupling of the cavity. The envelope
    lines depict the spectra convolved with a Gaussian function having a standard deviation of $\sigma=50$ cm$^{-1}$.}
\end{figure}

\begin{figure}[h!]
    \begin{tabular*}{1\textwidth}{@{\extracolsep{\fill}}ccc}
    
    \includegraphics[width=0.32\columnwidth]{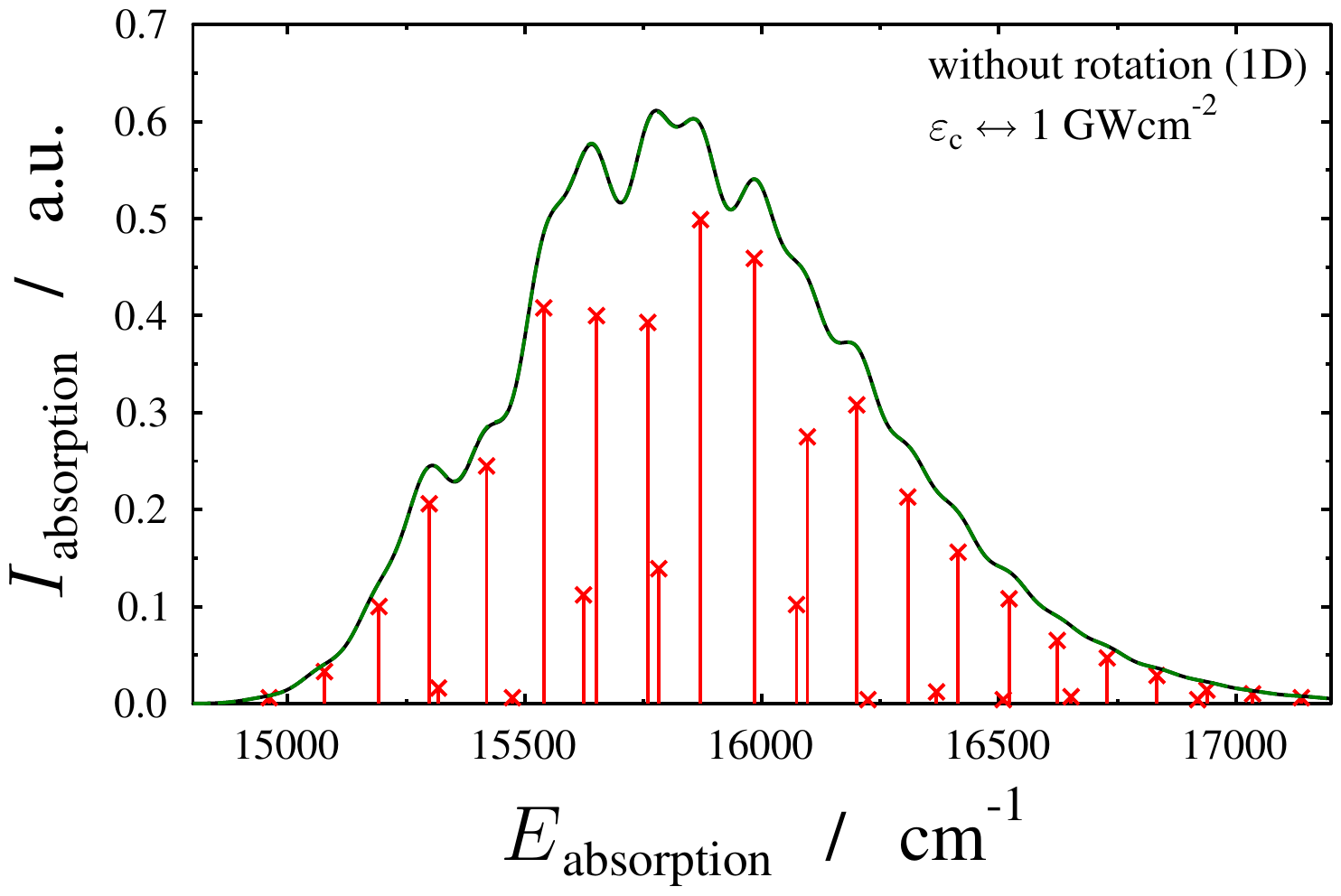} & \includegraphics[width=0.32\columnwidth]{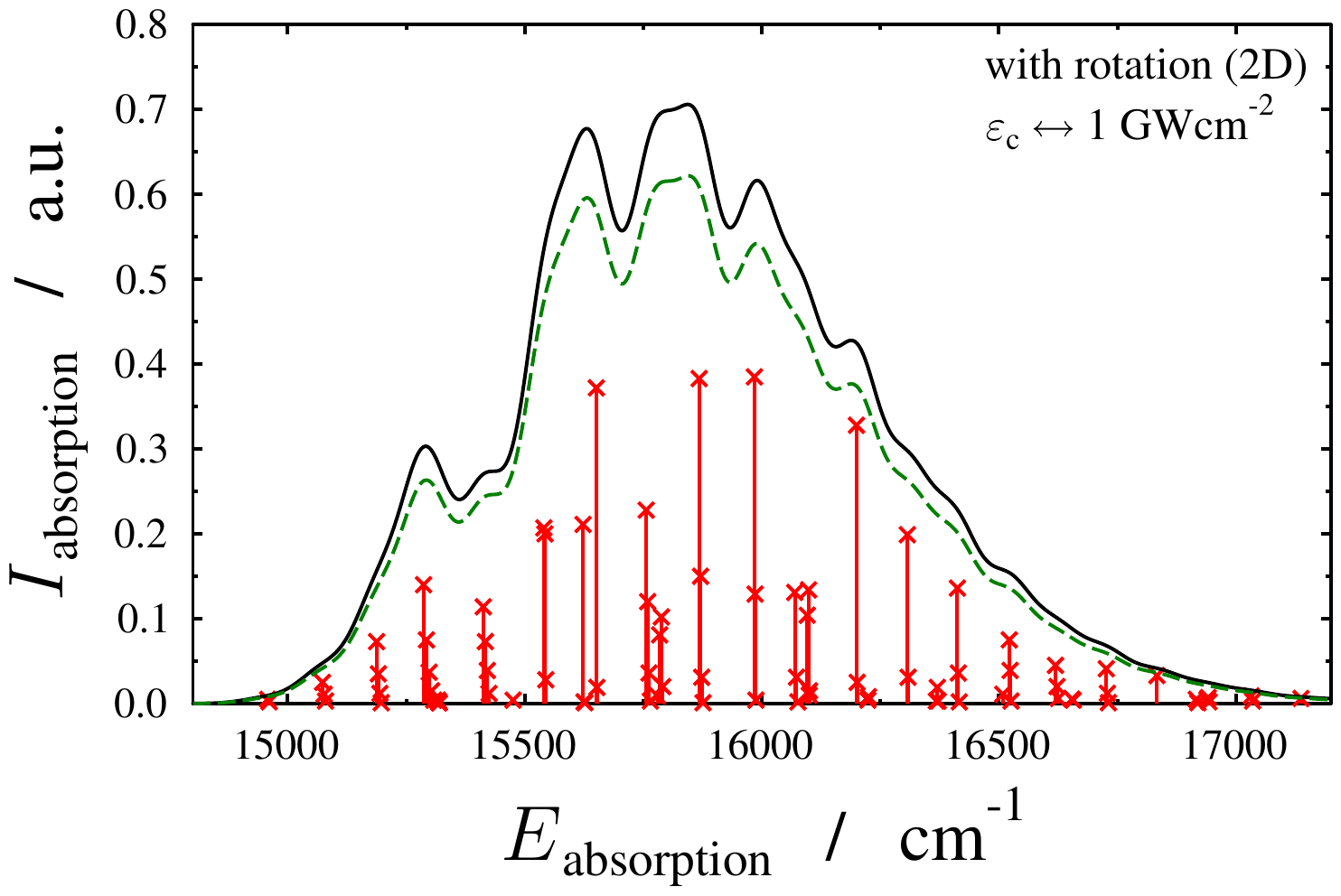} & \includegraphics[width=0.32\columnwidth]{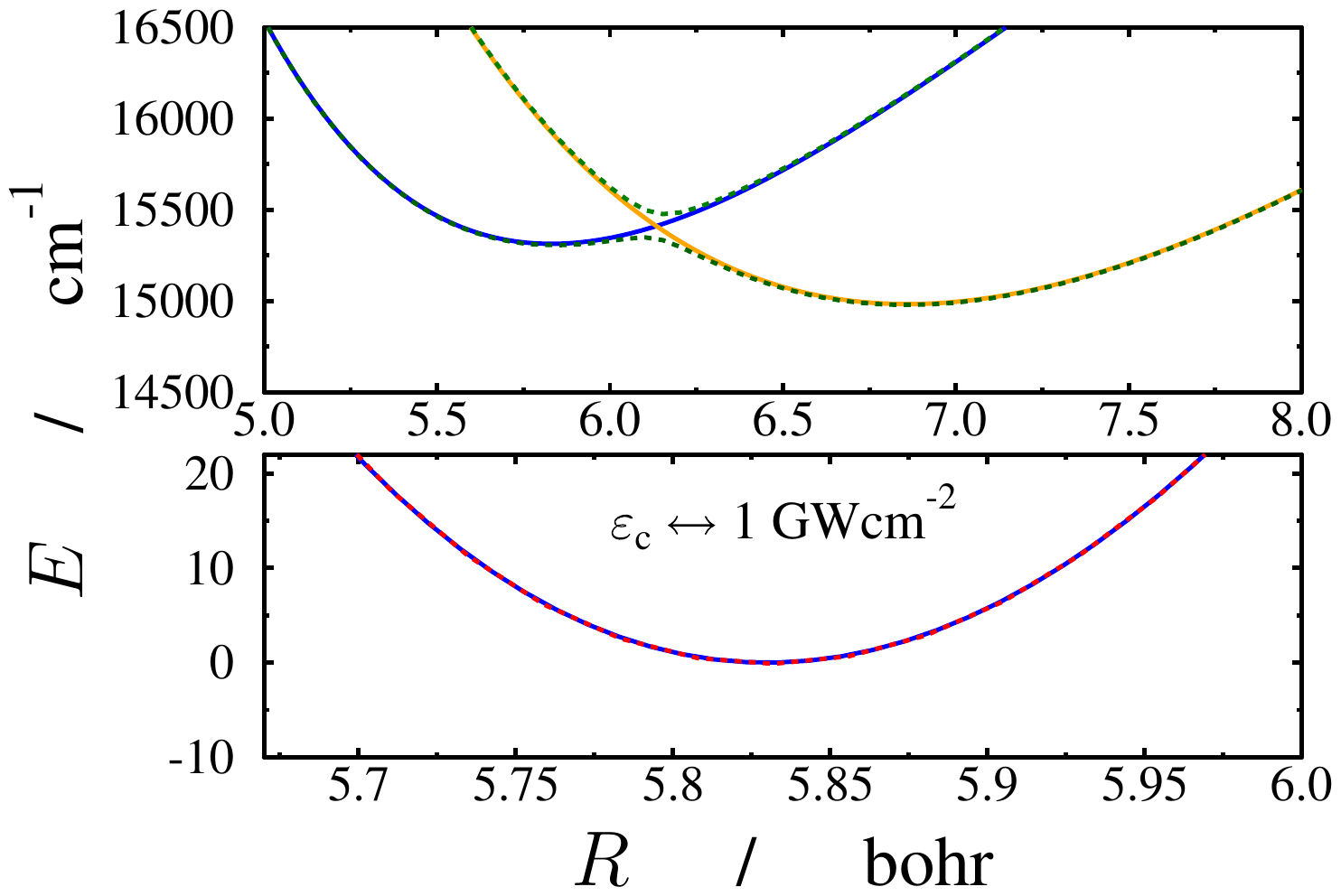}\tabularnewline
    
    \includegraphics[width=0.32\columnwidth]{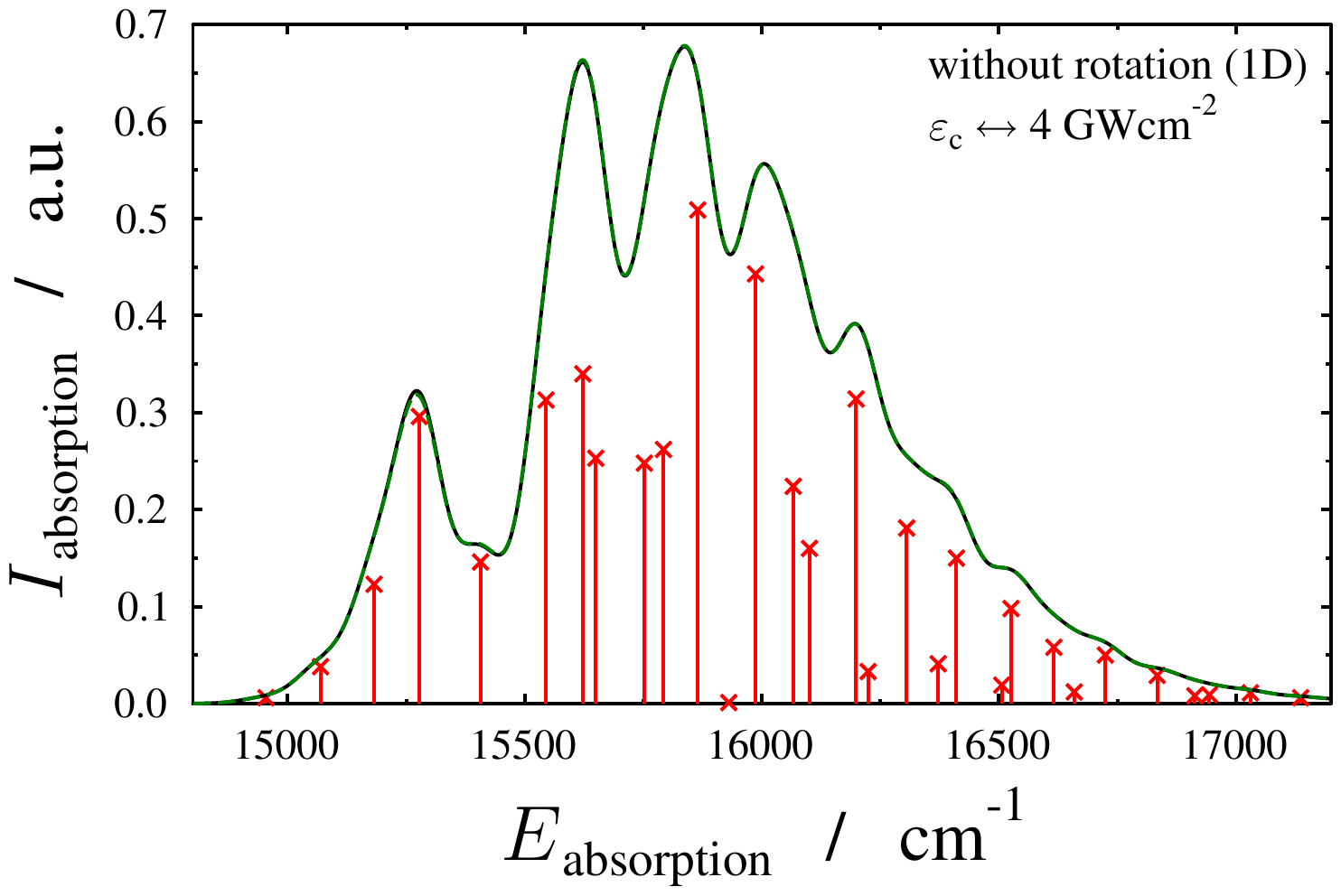} & \includegraphics[width=0.32\columnwidth]{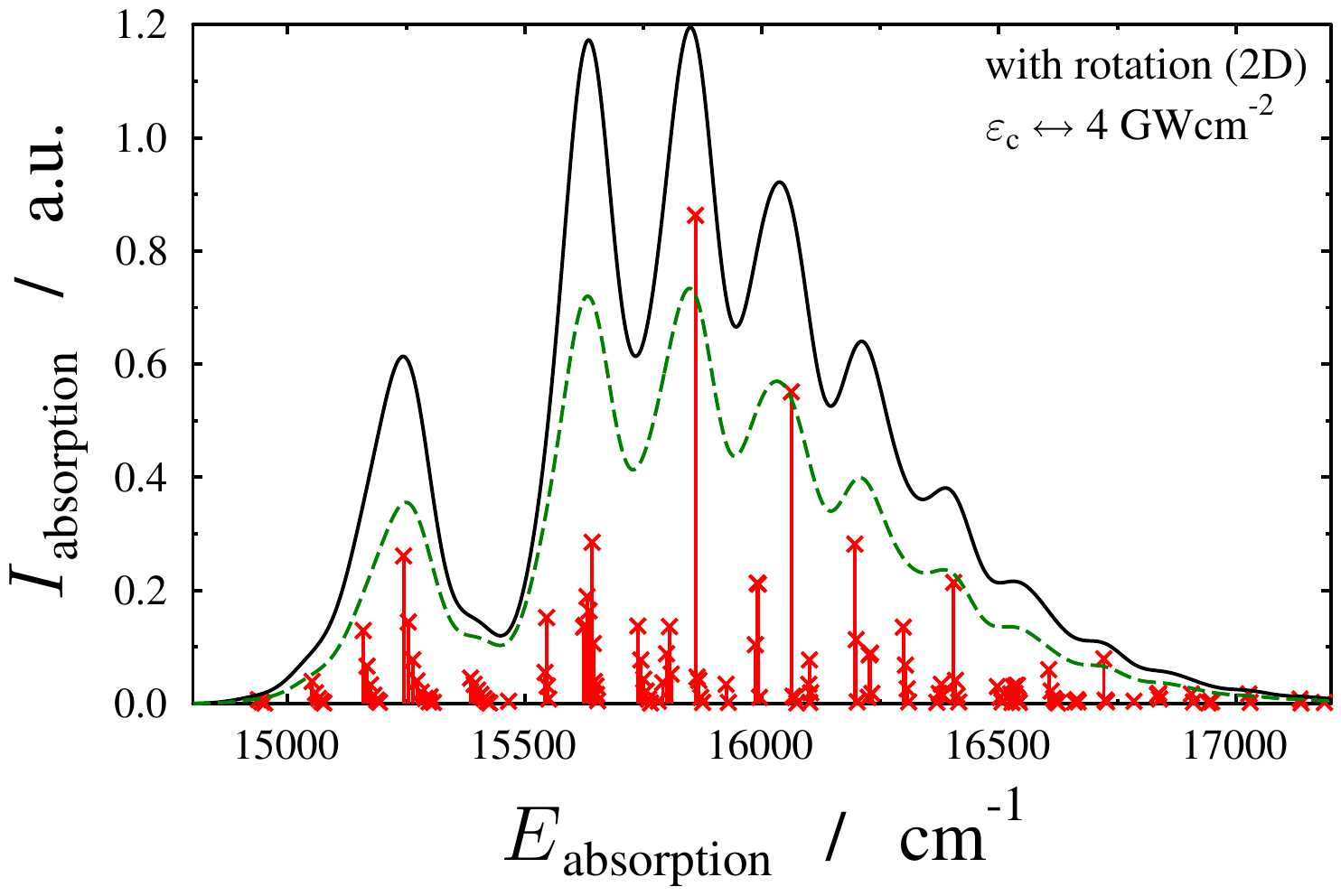} & \includegraphics[width=0.32\columnwidth]{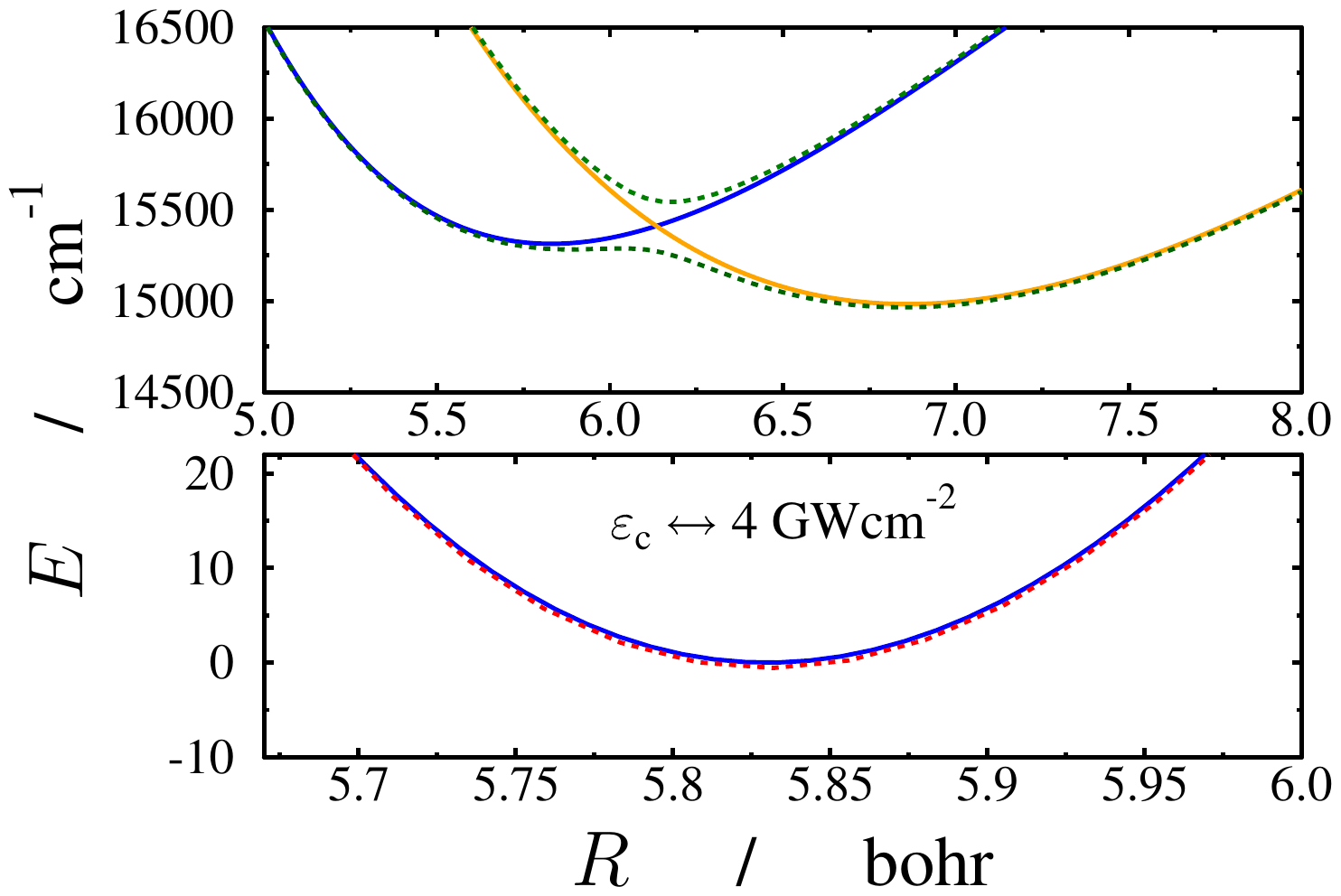}\tabularnewline

    \includegraphics[width=0.32\columnwidth]{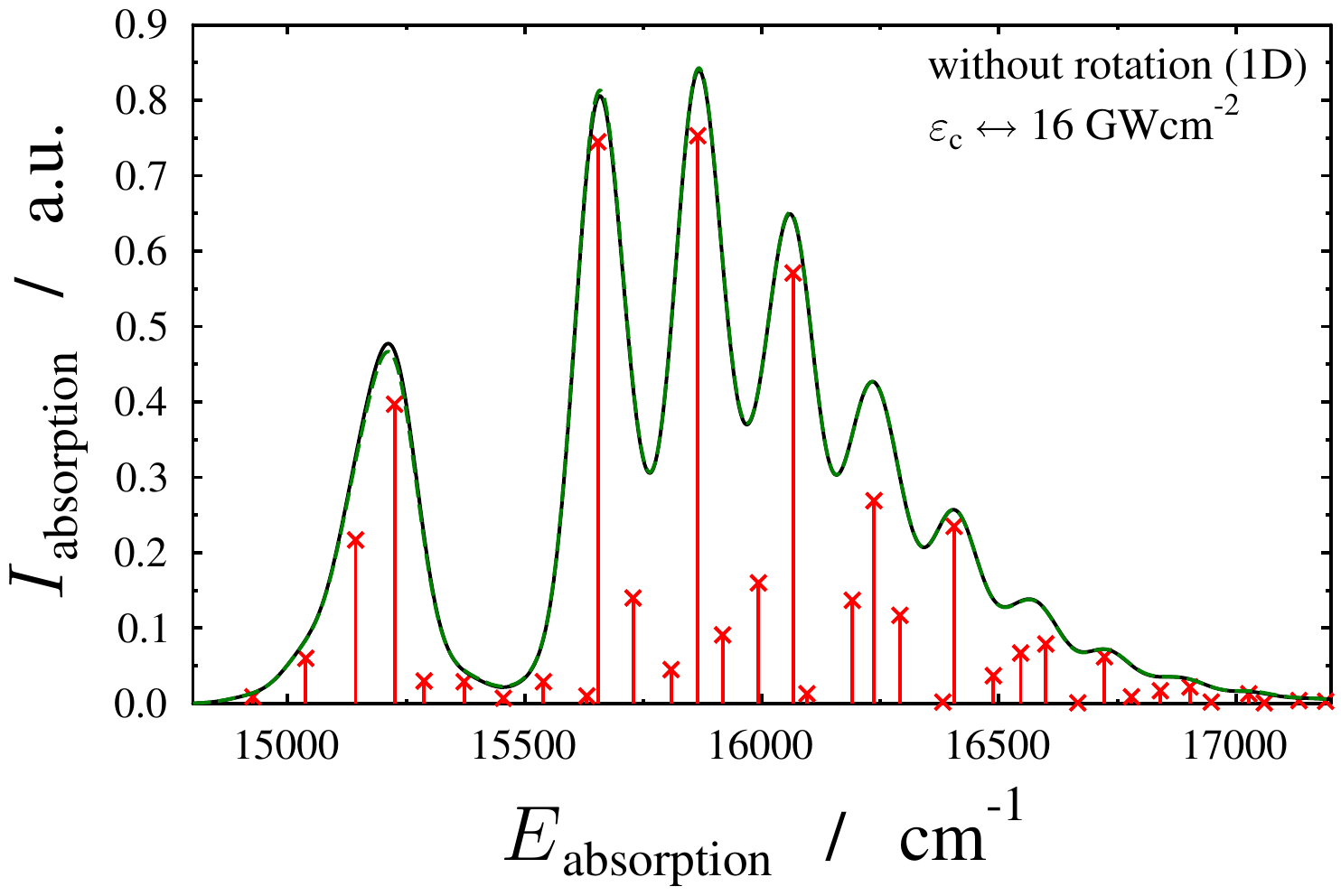} & \includegraphics[width=0.32\columnwidth]{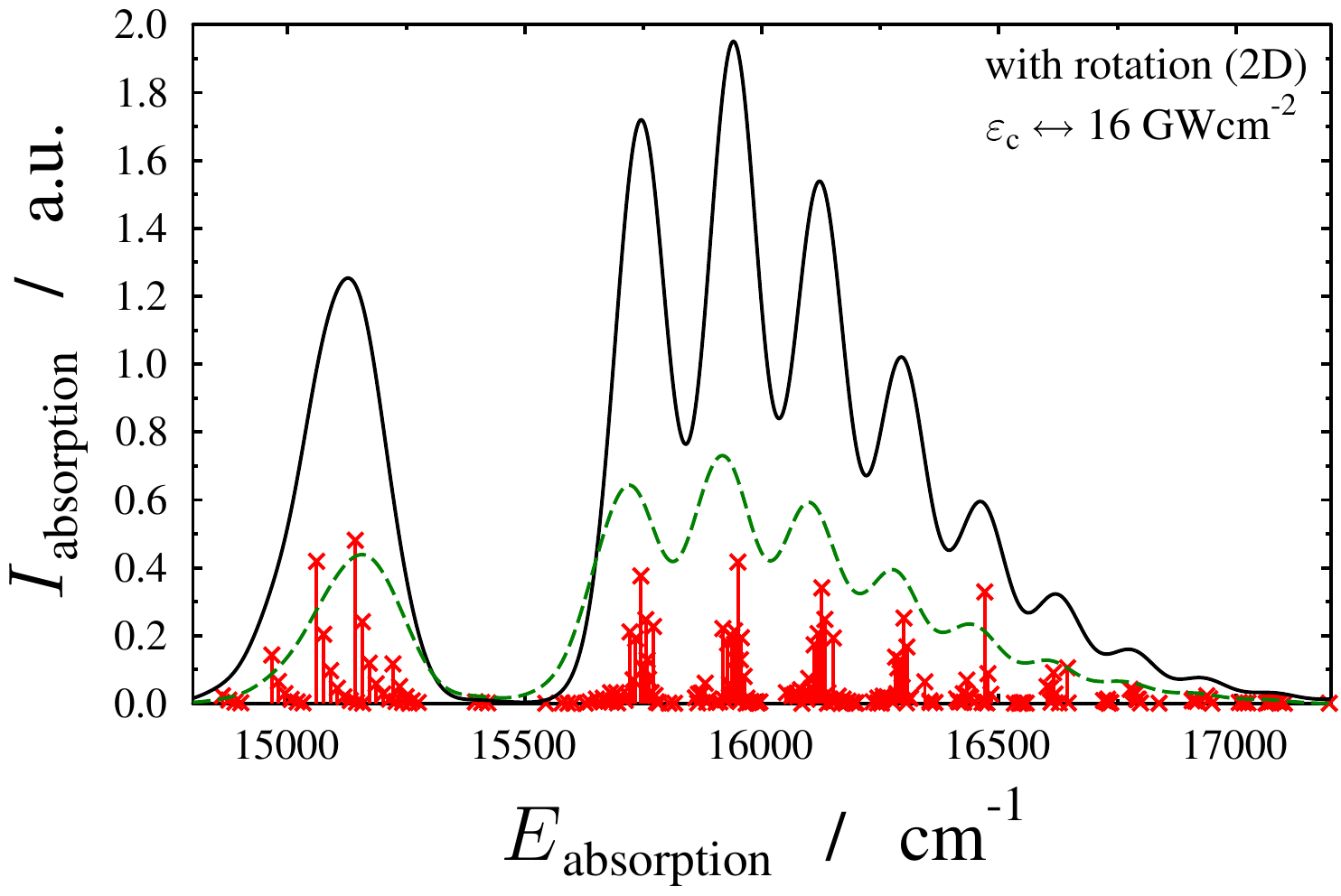} & \includegraphics[width=0.32\columnwidth]{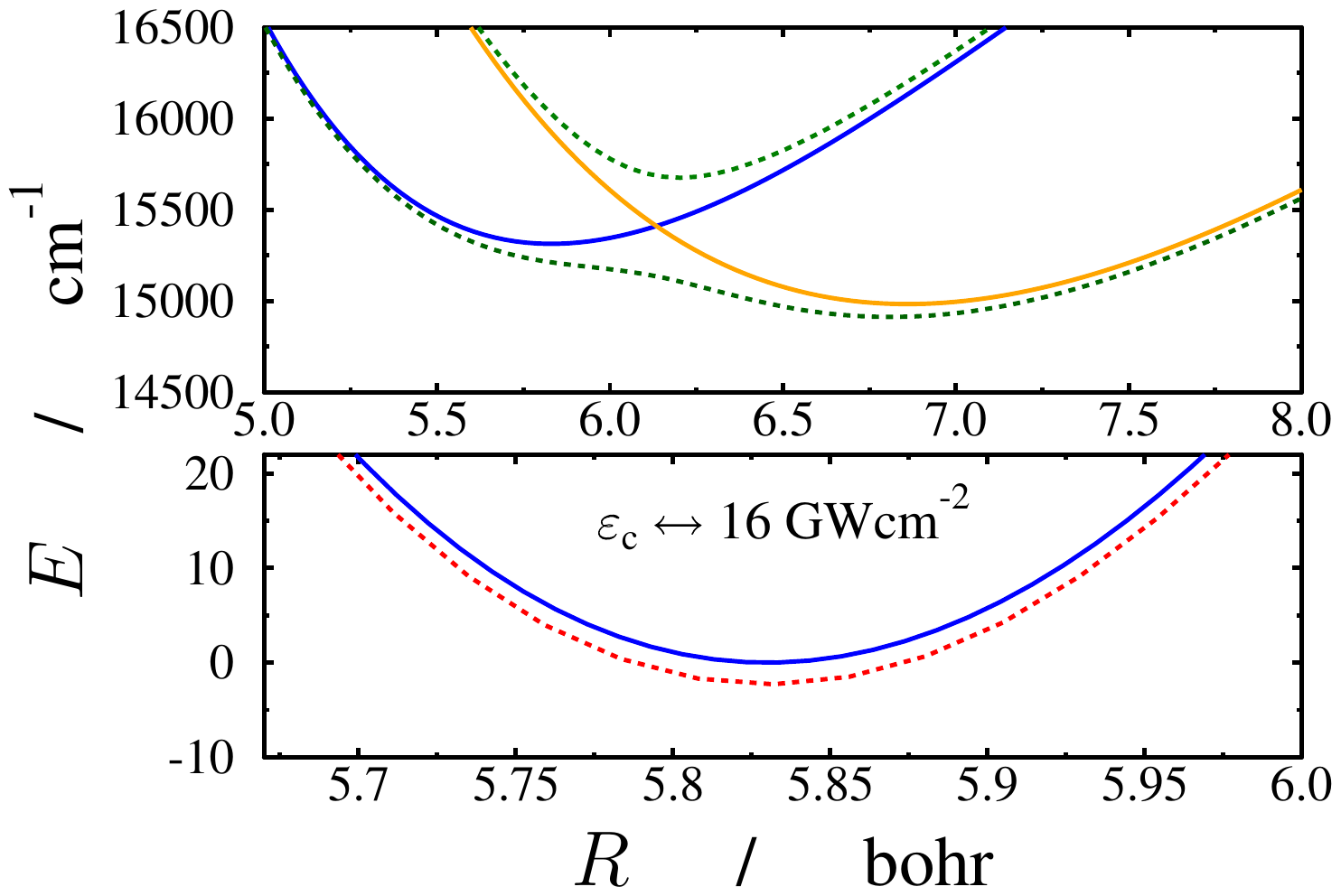}\tabularnewline
    
    \includegraphics[width=0.32\columnwidth]{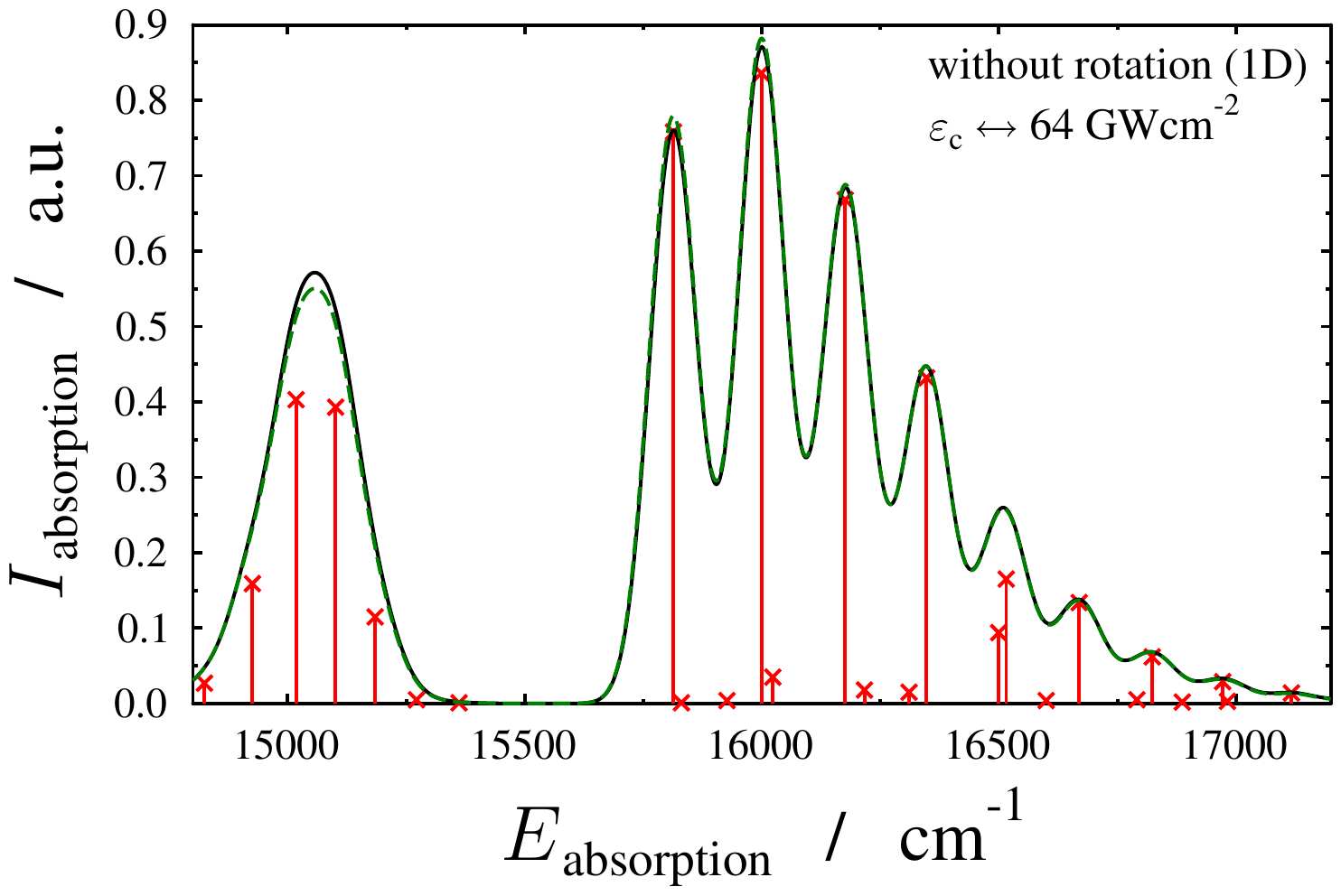} & \includegraphics[width=0.32\columnwidth]{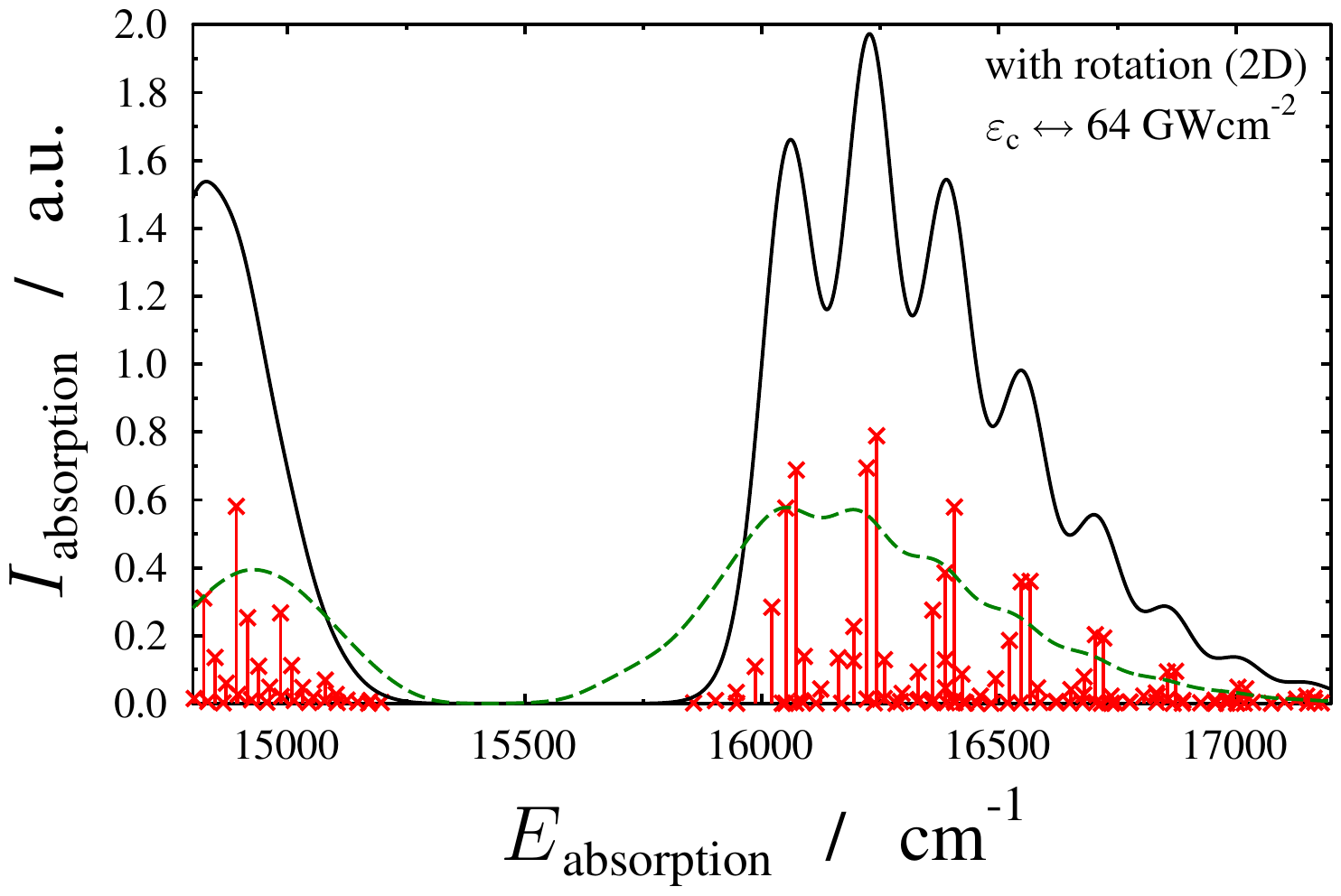} & \includegraphics[width=0.32\columnwidth]{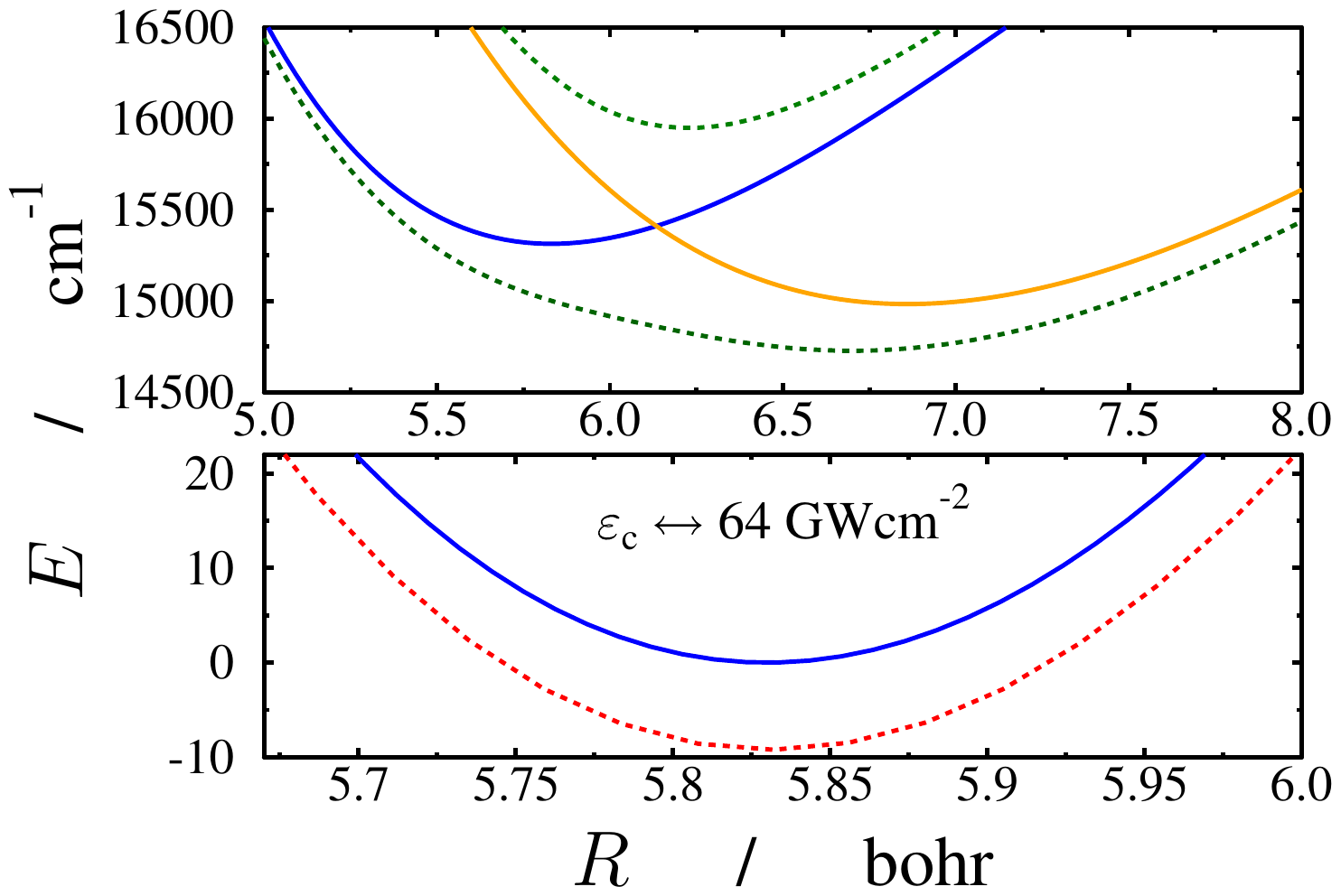}\tabularnewline
    \end{tabular*}%
    
    \caption{\label{FDS_weakToUScoupling}First two columns: field-dressed spectra obtained with different values of the light-matter coupling strengths for a cavity mode wavelength of $\lambda=653$ nm. Coupling strength values are indicated by the intensity of a classical light field giving a coupling strength equal to the one-photon coupling of the cavity. The envelope lines depict the spectra convoluted with a Gaussian function having a standard deviation of $\sigma=50$ cm$^{-1}$. The labels ``1D'' and ``2D'' stand for vibration only and rovibrational calculations, defined by using $J_{\rm max}=1$ and $J_{\rm max}=30$, respectively. Solid and dashed lines correspond to calculations including or excluding the off resonant couplings in the Hamiltonian, respectively. Third column: diabatic and adiabatic PECs at different light-matter coupling strength values.}
\end{figure}

\begin{figure}[h!]
    \includegraphics[width=0.45\columnwidth]{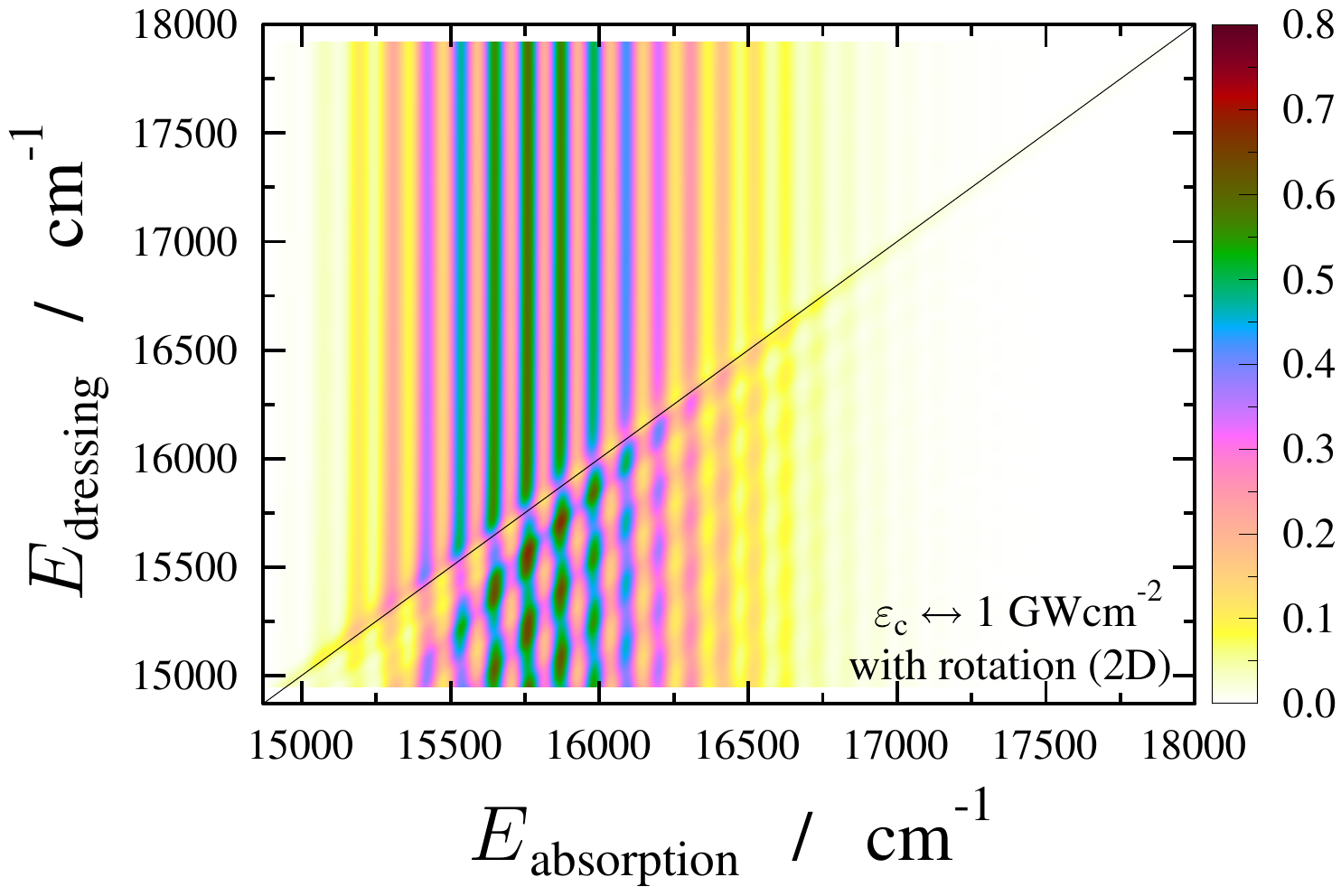}
    \includegraphics[width=0.45\columnwidth]{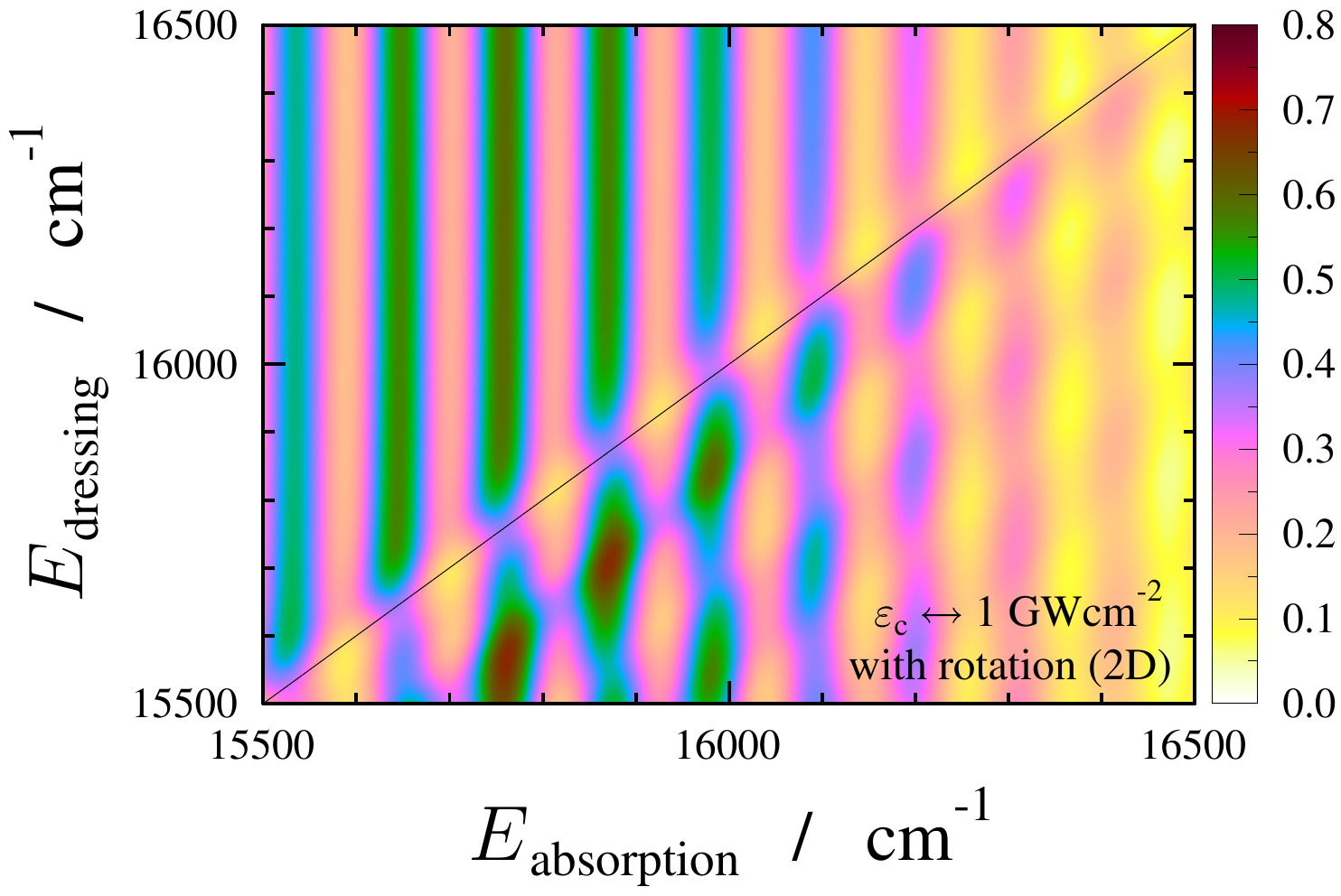}%
    \tabularnewline
    
    \includegraphics[width=0.45\columnwidth]{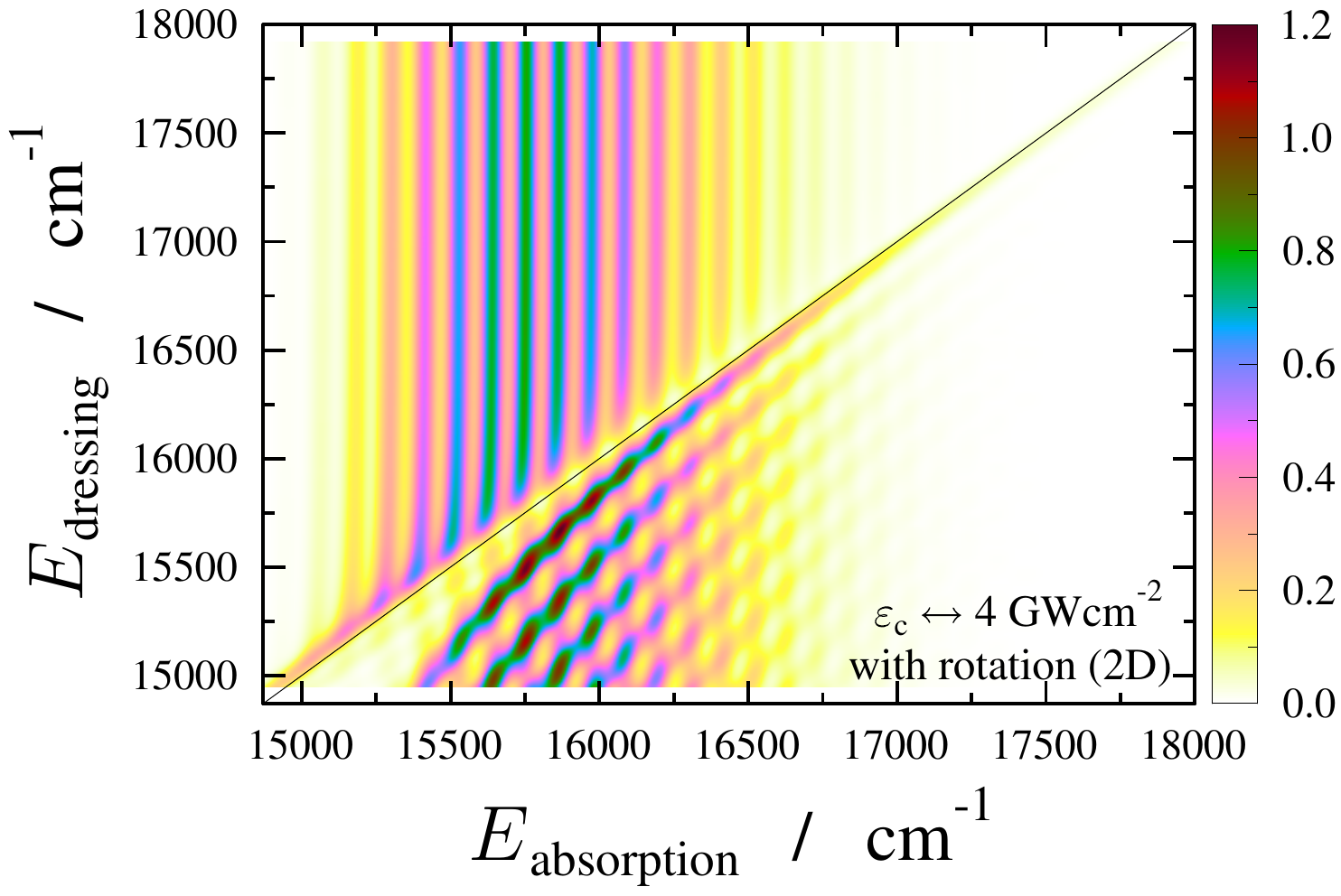}
    \includegraphics[width=0.45\columnwidth]{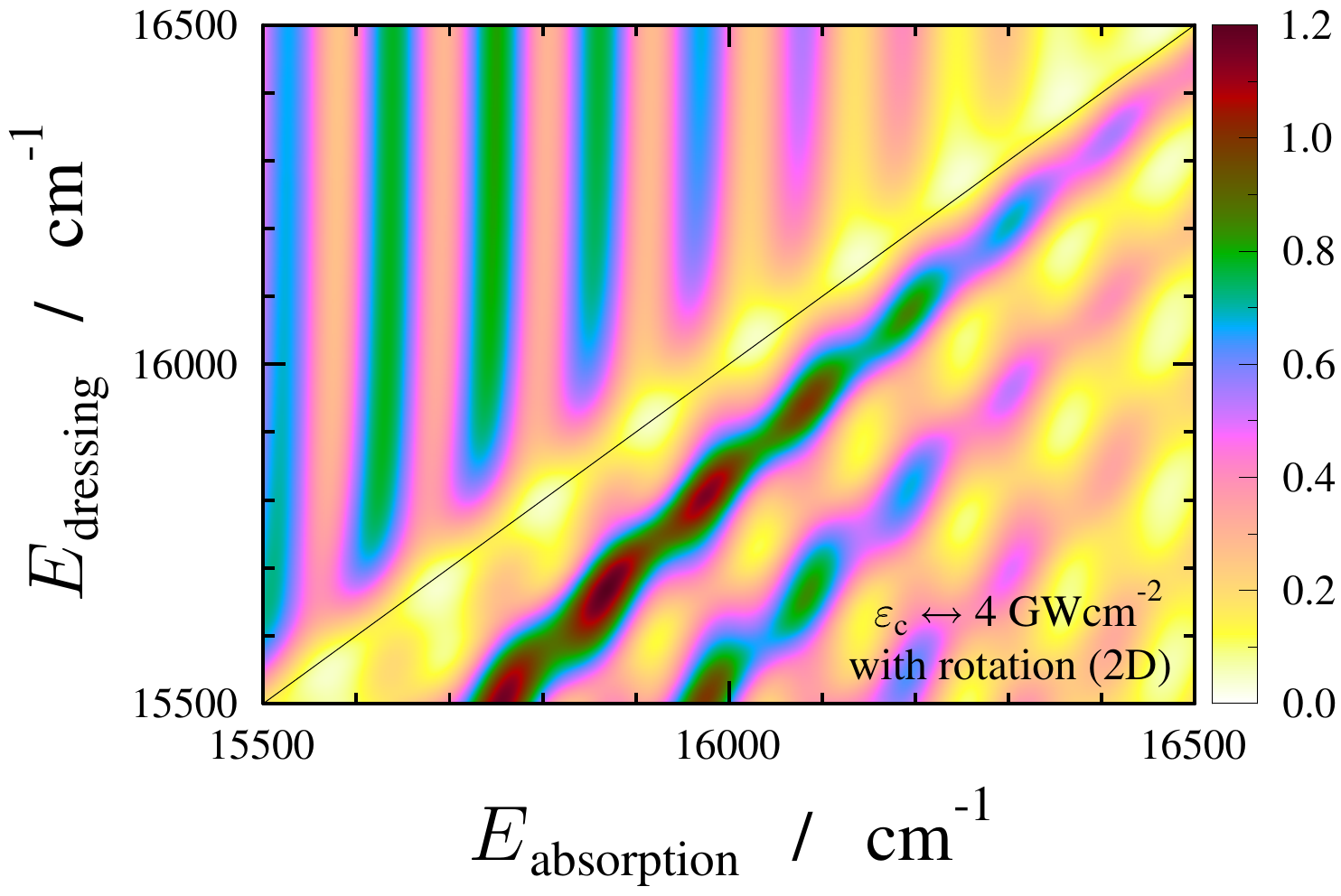}%
    \tabularnewline
    
    \includegraphics[width=0.45\columnwidth]{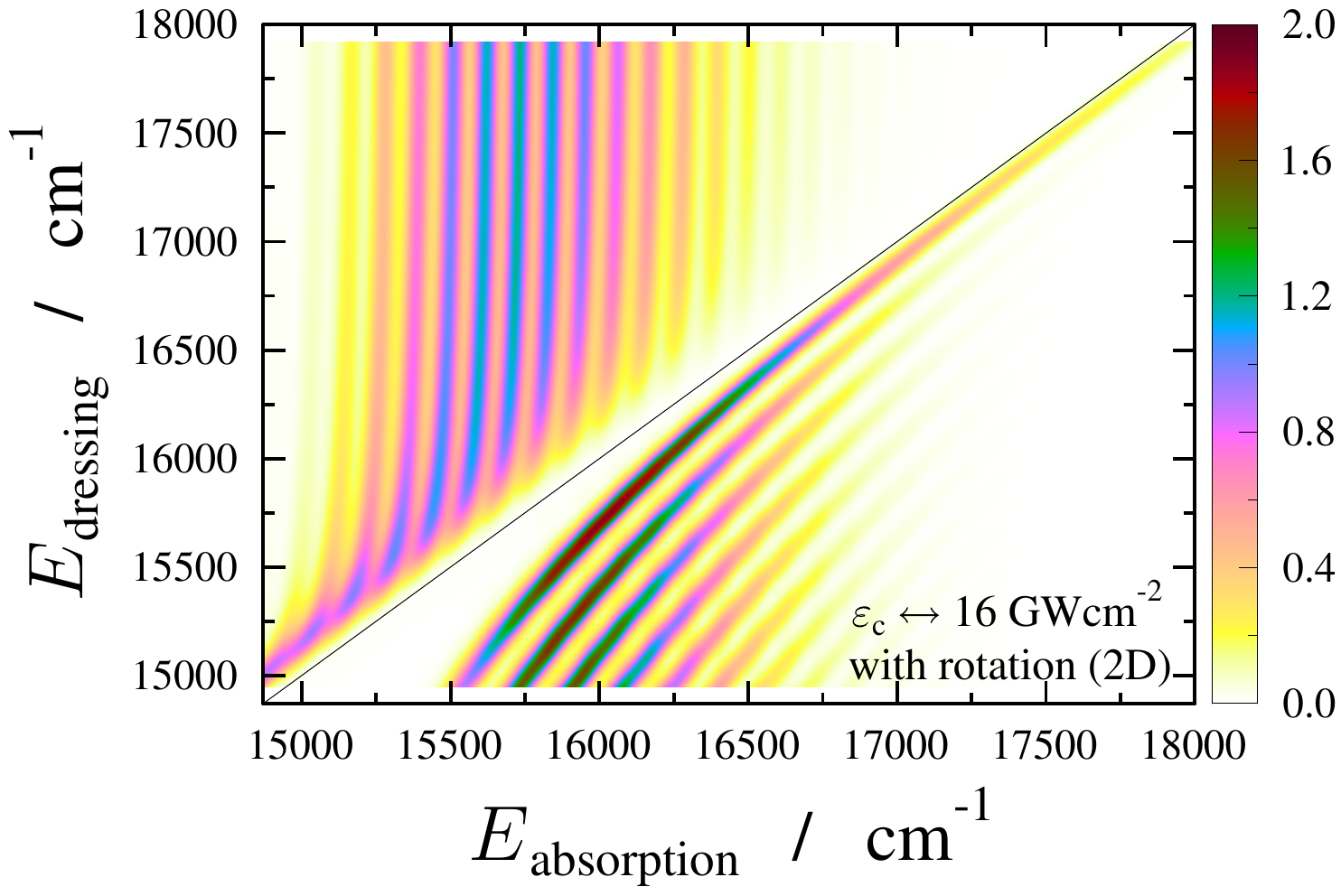}
    \includegraphics[width=0.45\columnwidth]{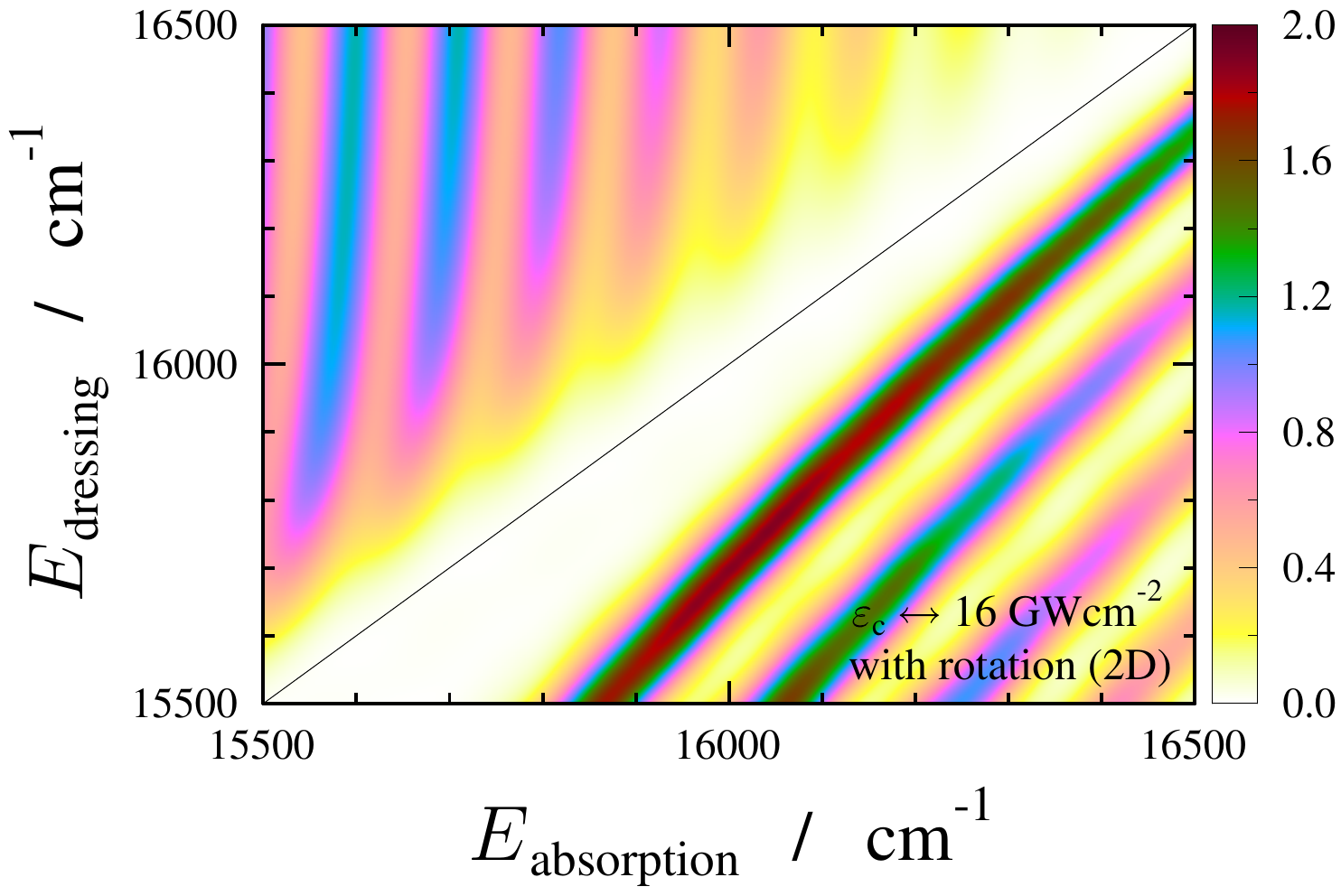}%
    \tabularnewline
    
    \caption{\label{W-I-D} Cavity mode wavenumber-dependence of the field-dressed spectrum obtained at three different coupling strengths, computed using Eq. (\ref{eq:transition_amplitude_between_cavity_FD_states}). Coupling strength values are indicated by the intensity of a classical light field giving a coupling strength equal to the one-photon coupling of the cavity. The spectrum is convoluted at each fixed cavity mode wavelength with a Gaussian function having $\sigma=30$ cm$^{-1}$.}
\end{figure}

    The PECs of Na$_2$ employed are given in Fig. \ref{PEC}, along with some important physical processes. The main results of this study are conveniently depicted in Figs. \ref{FDS_weakcoupling}--\ref{W-I-D}. Each figure will be discussed separately.

  \subsection{The field-dressed states}

\begin{figure}[h]
    \includegraphics[width=0.45\textwidth]{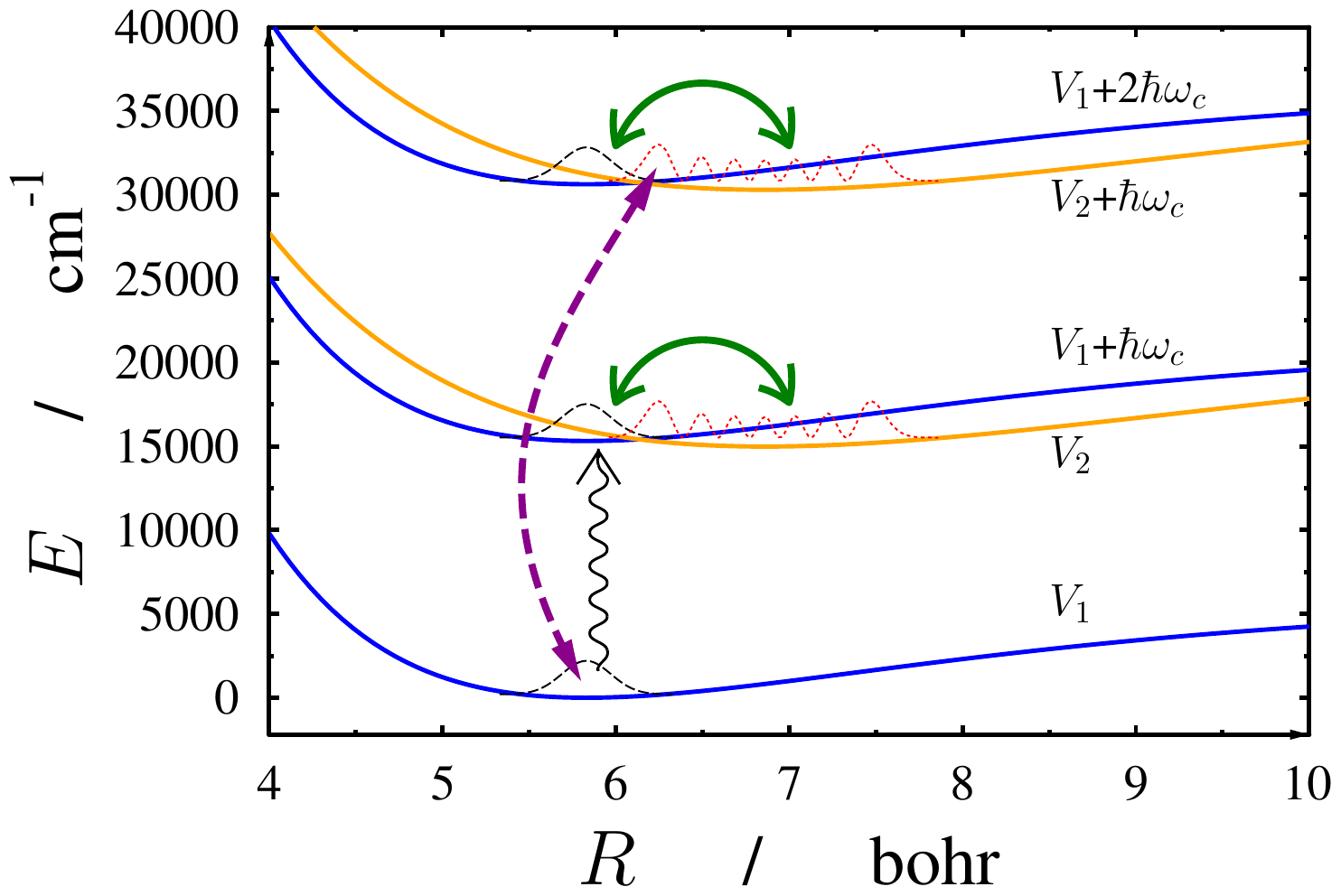}
    \includegraphics[width=0.45\textwidth]{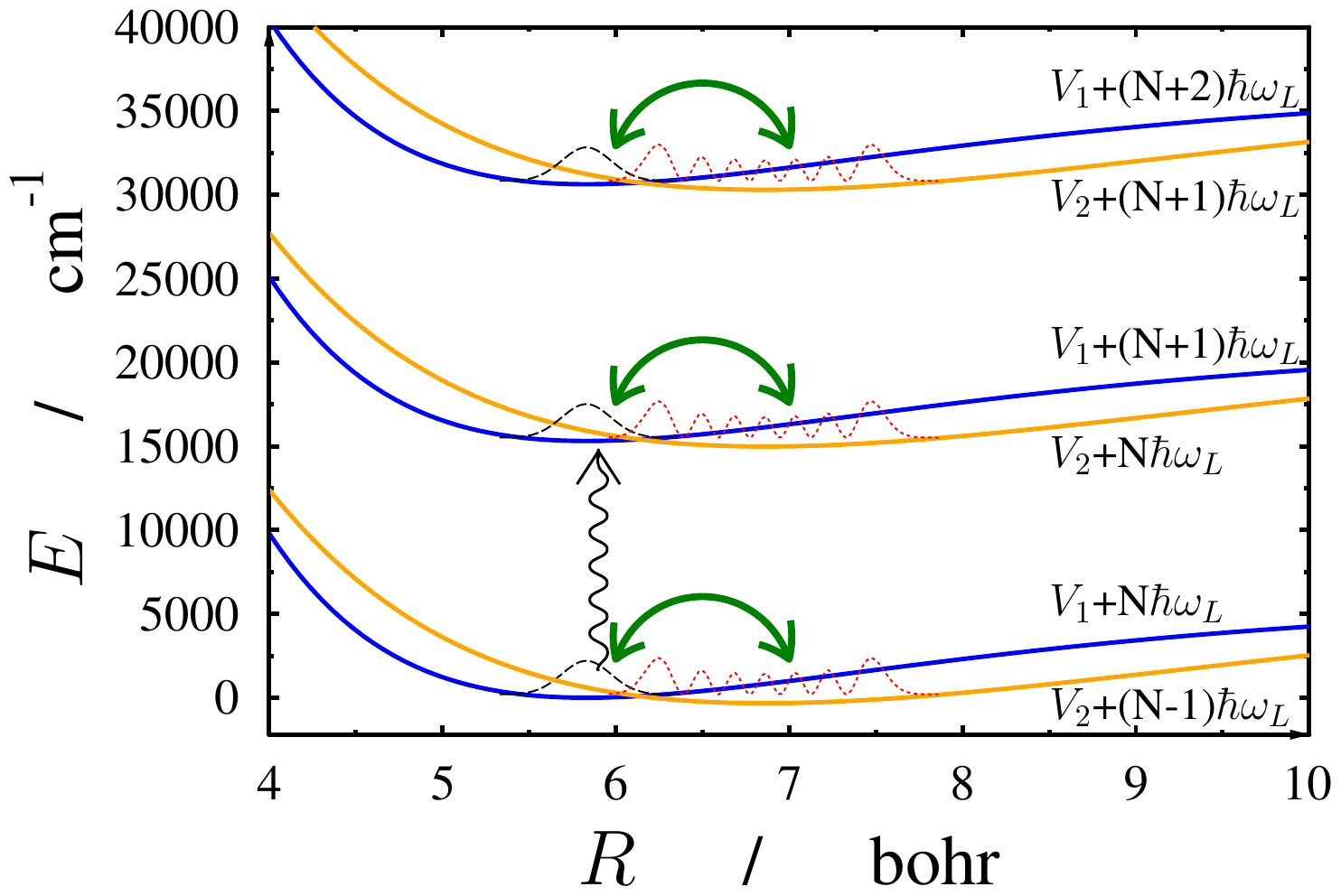}
    
    \caption{\label{PEC}Left: PECs of Na$_{2}$, dressed with different number of photons of the cavity field, obtained with a dressing-light wavelength of $\lambda=653$ nm. 
    Vibrational probability densities are drawn for states of $\vert1$
    $0$ $0\rangle\vert m\rangle$ type (dashed black lines on the $V_{1}(R)+m\hslash\omega_c$
    PECs), and for states of $\vert2$ $6$ $1\rangle\vert m\rangle$ type (dotted red lines on the $V_{2}(R)+m\hslash\omega_c$ PECs). Couplings
    induced by the cavity radiation field are indicated by the two double-headed
    arrows. The continuous green double-headed arrow represents $\vert1$
    $v$ $J\rangle\vert m\rangle\leftrightarrow\vert2$ $v'$ $J\pm1\rangle\vert m-1\rangle$
    type resonant couplings, while the dashed purple double-headed arrow
    represents $\vert1$ $v$ $J\rangle\vert m\rangle\leftrightarrow\vert2$
    $v'$ $J\pm 1\rangle\vert m+1\rangle$ type non-resonant couplings. Finally,
    the vertical brown wavy arrow indicates transitions between the two
    manifolds of field-dressed states, resulting form the absorption of
    a photon of the weak probe pulse. Right: same as left panel, but for Na$_2$ dressed by laser light. }
\end{figure}

    The left panel in Fig. \ref{PEC} shows PECs of Na$_{2}$ dressed with different number of photons in the cavity radiation field, as well as vibrational probability densities for direct-product states of the $\vert jvJ\rangle\vert N\rangle$ form. As apparent from Eq. (\ref{eq:CavityHamiltonian_detailed})
    and illustrated by the double-headed arrows in Fig.\ref{PEC}, light-matter
    interaction can give rise to resonant $\vert1$ $v$ $J\rangle\vert N\rangle\leftrightarrow\vert2$
    $v'$ $J\pm1\rangle\vert N-1\rangle$ and non-resonant $\vert1$ $v$
    $J\rangle\vert N\rangle\leftrightarrow\vert2$ $v'$ $J\pm1\rangle\vert N+1\rangle$
    type couplings, which lead to the formation of field-dressed states,
    see Eq. (\ref{eq:FieldDressedStates_small_photon_number}). The terms
    ``resonant'' and ``non-resonant''
    indicate whether the direct-product states that are coupled
    are close in energy or not, see Fig. \ref{PEC}. Naturally, resonant
    couplings are much more efficient in mixing the direct-product states
    than non-resonant couplings.
    For comparison, the right panel of Fig. \ref{PEC} shows the light-dressed PECs of Na$_{2}$ in a laser field \cite{LICI_in_spectrum_Szidarovszky_JPCL_2018}. Because nonresonant couplings are omitted in the usual Floquet description \cite{Floquet}  of laser light-dressed molecules, these couplings are not shown in the right panel of Fig. \ref{PEC}. It is clear from Fig. \ref{PEC} that the absorption spectrum of field-dressed molecules should be considerably different for the cavity-dressed and laser-dressed cases. The most significant difference is that while in the cavity the ground state is primarily a field-free eigenstate in vacuum, which is only deformed at relatively large coupling strengths through nonresonant couplings, the laser light-dressed state correlating to the field-free ground state contains a mixture of field-free eigenstates due to the strong resonant coupling in this case. 
    
\subsection{Spectra in the weak coupling regime}

    Figure \ref{FDS_weakcoupling} shows field-dressed spectra obtained with different values of the $\varepsilon _c = \sqrt{\hslash\omega_c/(\epsilon_{0}V)}$ cavity one-photon field  strength in the weak-coupling regime for a cavity-mode wavelength of $\lambda=653$ nm. Although the light-matter coupling strength and the cavity-mode wavelength are not completely independent in a cavity (see Eq. (\ref{eq:quantized_A})), we treat them as independent parameters. This can be rationalized partially by Eq. (\ref{eq:quantized_A}), which shows that the coupling strength could be changed independently by changing the cavity volume, while keeping the cavity length responsible for the considered cavity radiation mode fixed.
    The spectra in Fig. \ref{FDS_weakcoupling} were computed assuming that the initial state is the
    ground state of the full system. The left panel of Fig. \ref{FDS_weakcoupling} reflects features similar to those observed
    in the spectrum of Na$_{2}$ dressed by laser fields \cite{LICI_in_spectrum_Szidarovszky_JPCL_2018}.
    With increasing light-matter coupling, the overall intensity of the
    spectrum slightly increases at almost all wavenumbers, with some
    shoulder features becoming more pronounced in the spectrum envelope. In terms of spectroscopic nomenclature, such a phenomenon can be understood as an intensity-borrowing effect \cite{Cederbaum_multimode,vibronic_coupling_model_Cederbaum_AnnRevPhysChem_2004,CI_spectroscopyYarkony_AnnRevPhysChem_2012}, which arises from the field-induced couplings between field-free states.
    
    On the other hand, the coupling strength dependence of the spectrum envelope is completely absent if the spectra in Fig. \ref{FDS_weakcoupling} are generated from computations in which the rotational motion is restricted by setting $J_{\rm max}=1$. Such a rotationally-restricted model inherently lacks any signatures of a LICI, whose formation requires at least two nuclear degrees of freedom. Therefore, in terms of the adiabatic representation, the intensity borrowing effect visible in Fig. \ref{FDS_weakcoupling} can be attributed to the nonadiabatic couplings of a LICI created by the quantized cavity radiation field.
    
    Inspecting the individual transition lines in the spectra reveals that increasing
    the light-matter coupling strength can result in the splitting
    of existing peaks and the appearance of additional peaks, as shown
    in the two panels on the right side of Fig. \ref{FDS_weakcoupling}. The upper right panel of Fig. \ref{FDS_weakcoupling}
    shows the progression of three peaks, corresponding to transitions
    from the initial state (essentially the $\vert1$ $0$ $0\rangle\vert0\rangle$
    state) to field-dressed states composed primarily of the $\vert2$
    $7$ $1\rangle\vert0\rangle$, $\vert2$ $7$ $3\rangle\vert0\rangle$,
    and $\vert2$ $7$ $5\rangle\vert0\rangle$ states, with $\vert1$
    $v$ $J\rangle\vert1\rangle$-type states ($J$ even) contributing
    as well. With increasing light-matter coupling strength these transitions
    are red shifted, and they can be interpreted as originating from the
    field-free transition $\vert1$ $0$ $0\rangle\rightarrow\vert2$
    $7$ $1\rangle$, which is split due to the mixing of $\vert2$ $7$
    $1\rangle$ with other states through the light-matter coupling with
    the cavity mode. The lower right panel of Fig. \ref{FDS_weakcoupling} shows
    the progression of three peaks, which do not arise from the splitting
    of an existing field-free peak, but appear as new peaks. These transitions
    are blue shifted with increasing light-matter coupling strength, and
    they occur between the initial state (essentially the $\vert1$ $0$
    $0\rangle\vert0\rangle$ state) and field-dressed states composed
    primarily of the $\vert1$ $3$ $0\rangle\vert1\rangle$, $\vert1$
    $3$ $2\rangle\vert1\rangle$, and $\vert1$ $3$ $4\rangle\vert1\rangle$ states. Such transitions are forbidden in the zero light-matter coupling
    limit; however, they become visible as the light-matter coupling with
    the cavity mode contaminates the $\vert1$ $3$ $J\rangle\vert1\rangle$
    states with $\vert2$ $v$ $1\rangle\vert0\rangle$-type states, to
    which the initial state has allowed transitions.

\subsection{Spectra in the strong and ultrastrong coupling regimes}

    In Fig. \ref{FDS_weakToUScoupling} field-dressed spectra obtained with
    light-matter coupling strengths ranging from the weak to the ultrastrong coupling regimes are shown for a cavity-mode wavelength of $\lambda=653$ nm. The spectra labeled ``1D'' in Fig. \ref{FDS_weakToUScoupling} were obtained with a model having restricted rotational motion ($J_{\rm max}=1$). The results labeled ``2D'' fully account for rotations as well as vibrations; therefore, incorporate the effects of a LICI on the spectrum.
    By comparing the ``1D'' and ``2D'' spectrum envelopes in Fig. \ref{FDS_weakToUScoupling}, it is apparent that there is a significant increase in absorption for the ``2D'' case. As discussed in the next subsection, this is primarily due to nonresonant light-matter couplings between $\vert1$ $0$ $J\rangle\vert0\rangle$ and $\vert2$ $v'$ $J\pm 1\rangle\vert1\rangle$-type states and partially due to the intensity borrowing effect induced by the nonadiabatic couplings of the LICI. 
    
    Figure \ref{FDS_weakToUScoupling} also shows field-dressed PECs in the diabatic and adiabatic representations. 
    As the coupling strength increases, two separate polariton surfaces are formed, and the absorption spectrum splits into two distinct groups of peaks, corresponding to transitions onto the two polariton states. At the largest coupling strength, a slight modification of the ground-state PEC can also be seen, indicating that the ultrastrong coupling regime has been reached.

\subsection{Impact of nonresonant coupling}

    Interestingly, nonresonant couplings seem to have an impact on the spectrum at much smaller coupling strengths than those required for a significant modification of the ground-state PEC. The spectrum envelopes depicted with dotted and continuous lines in Fig. \ref{FDS_weakToUScoupling} indicate whether spectra were computed by using the 
    \begin{equation}
    \hat{H}_{\rm 3x3}=
    \begin{bmatrix}\hat{T}+V_{1}(R) & 0 & 0  \\
    0 & \hat{T}+V_{2}(R) & g_{21}(R,\theta)\sqrt{2}  \\
    0 & g_{12}(R,\theta)\sqrt{2} & \hat{T}+V_{1}(R)+\hslash\omega_c 
    \end{bmatrix},\label{eq:CavityHamiltonian_detailed_3x3}
    \end{equation}
    upper left three-by-three block of the Hamiltonian in Eq. (\ref{eq:CavityHamiltonian_detailed}) or a six-by-six block, respectively. The deviation between these two types of spectra represents the effects of nonresonant couplings, because the $\hat{T}+V_{2}(R)+\hslash\omega_c$ term and its couplings with $\hat{T}+V_{1}(R)$ are present in Eq. \ref{eq:CavityHamiltonian_detailed}, but absent in Eq. \ref{eq:CavityHamiltonian_detailed_3x3}. 
    
    The plots in Fig. \ref{FDS_weakToUScoupling} clearly demonstrate that in the vibration-only ``1D'' case nonresonant couplings have no visible impact on the spectrum; however, for the ``2D'' case, in which rotations are accounted for, nonresonant couplings lead to a visible increase in the absorption signal even at the lowest coupling strengths shown. The physical origin of the increase in absorption is the contamination of the $\vert1$ $0$ $0\rangle\vert0\rangle$ ground state with the $\vert1$ $0$ $2\rangle\vert0\rangle$, $\vert1$ $0$ $4\rangle\vert0\rangle$, etc. states, which allows for transitions onto the $J=3,5,...$ components of the rovibronic states in the excited polariton manifold. These results indicate that the effects of nonresonant couplings can not be described in a vibration only model, and if one wishes to obtain meaningful simulation results for coupling strengths reaching or exceeding those shown in Fig. \ref{FDS_weakToUScoupling}, it is necessary to properly account for molecular rotations.

\subsection{Cavity-mode wavelength dependence of the spectrum}

    Figure \ref{W-I-D} shows the cavity-mode wavenumber dependence of the
    field-dressed spectrum obtained at the $\varepsilon _c=\sqrt{\hslash\omega_c/(\epsilon_{0}V)}$ cavity one-photon field strengths of 0.844$\cdot 10^{-4}$, 1.688$\cdot 10^{-4}$, and 3.376$\cdot 10^{-4}$ atomic units, corresponding to classical field intensities of 1, 4, and 16 GWcm$^{-2}$, respectively.
    It can be concluded from Fig. \ref{W-I-D} that, as expected, the field-dressed spectrum changes with the cavity-mode wavelength. Furthermore, the cavity-mode wavelength dependent spectrum shows qualitative features considerably different from the dressing-field wavelength dependence of the spectrum when Na$_{2}$ is dressed by medium intensity laser fields \cite{LICI_in_spectrum_Szidarovszky_JPCL_2018}, as one might expect from Fig. \ref{PEC}.
    
    For all coupling strengths shown in Fig. \ref{W-I-D}, at large dressing-field photon energies, \textit{i.e.}, those exceeding 17000 cm$^{-1}$ or so, the spectra resemble the field-free spectrum, depicting around twenty lines corresponding to transitions to $\vert2$ $v$ $1\rangle\vert0\rangle$-type states. This is expected, because for such large photon energies, the $V_1(R)+\hslash \omega _c$ PEC crosses the $V_2(R)$ PEC at short internuclear distances, and the $V_2(R)$ PEC remains unperturbed in the Frank--Condon region. As the photon energy of the dressing field is lowered and the crossing of the $V_1(R)+\hslash \omega _c$ and $V_2(R)$ PECs approaches the Frank--Condon region, the spectrum becomes perturbed. 
    
    In the top row of Fig. \ref{W-I-D}, a decrease can be seen in the spectrum line intensities along diagonal lines in the plots, forming island-type features. Focusing on a specific vibrational state on $V_2(R)$, corresponding to a vertical line in the plots, a decrease in the spectrum intensity occurs when this vibrational state becomes resonant with one of the $\vert1$ $v$ $J\rangle\vert1\rangle$ states. Due to the resonance, a strong mixing occurs between the $\vert1$ $v$ $J\rangle\vert1\rangle$- and $\vert2$ $v'$ $J'\rangle\vert0\rangle$-type states, which leads to a decrease of the transition amplitude from the ground state. Nonetheless, when the mixing of the states is not as efficient as in the resonant case, \textit{i.e.}, at the island-type features on the plots, an increase can be seen in the spectrum intensities with respect to the field-free case.
    
    As depicted in the middle and bottom rows of Fig. \ref{W-I-D}, when the coupling strength is increased, the picture of a ``perturbed spectrum'' gradually changes into the picture of two distinct spectra corresponding to the two polariton surfaces, in accordance with Fig. \ref{FDS_weakToUScoupling}. The dressing-field wavenumber dependence of the spectrum in the bottom row of Fig. \ref{W-I-D} can easily be understood in terms of the wavenumber dependence of the polariton surfaces depicted in the rightmost column of Fig. \ref{FDS_weakToUScoupling}.

\section{Summary and conclusions}

    We investigated the rovibronic spectrum of homonuclear diatomic molecules dressed by the quantized radiation field of an optical cavity. Formation of light-induced conical intersections induced by the quantized radiation field is shown for the first time by identifying the robust light-induced nonadiabatic effects in the spectrum. The coupling strength and the cavity mode wavelength dependence of the field-dressed spectrum was also investigated from the weak to the ultrastrong coupling regimes. Formation of polariton states in the strong coupling regime was demonstrated, and its was shown how nonresonant couplings lead to an increased absorption in the field-dressed spectrum even before the ultrastrong coupling regime is reached. The numerical results demonstrate that the additional degree of freedom (which is the rotation in the present diatomic situation) plays a crucial role in the appropriate description of the light-induced nonadiabatic processes as well as in the efficiency of nonresonant couplings. Therefore, for physical scenarios when diatomic molecular rotations can proceed in the cavity, properly accounting for the rotational degrees of freedom is mandatory for obtaining reliable simulation results. 

    We hope that our findings will stimulate photochemical cavity experiments, and also the extension of the theory for the proper description of polyatomic molecules. It did not escape our attention that there is much potential in studying light-induced conical intersections in polyatomic molecules in cavity without rotations as there are many nuclear degrees of freedom to form such intersections which can also be used to selectively manipulate certain chemical and physical properties.

\section{Acknowledgement}

    This research was supported by the EU-funded Hungarian grant EFOP-3.6.2-16-2017-00005 and by the Deutsche Forschungsgemeinschaft (Project ID CE10/50-3). The authors are grateful to NKFIH for support (Grant No. PD124623, K119658 and K128396). The authors thank P\'eter Domokos for the fruitful discussions.

\bibliography{Na2_in_cavity}

\end{document}